\documentclass[]{jfm}

\graphicspath{{figures/}} 
\usepackage{graphicx} 
\usepackage{natbib} 
\usepackage{hyperref} 
\usepackage{newtxtext,newtxmath}

\newcommand{\vect}[1]{\boldsymbol{#1}} 
\newcommand{\aver}[1]{ \! \left\langle {#1} \right \rangle \!}
\newcommand{\paver}[1]{ \! \left\langle {#1} \right \rangle_p \!}
\newcommand{\thaver}[1]{ \! \left\langle {#1} \right \rangle_\theta \!}
\title[Drag reduction and subcritical turbulence in controlled pipe flows]{Drag reduction and subcritical turbulence in controlled pipe flows}

\author[]{Emanuele Gallorini\aff{1,}\aff{2}\corresp{\email{emanuele.gallorini@polimi.it}}, Daniele~Massaro\aff{3}, Philipp~Schlatter\aff{4} and Maurizio~Quadrio\aff{1}}
\affiliation{
\aff{1} Dipartimento di Scienze e Tecnologie Aerospaziali, Politecnico di Milano, via La Masa 34, 20156 Milano, Italy
\aff{2} Univ. Lille, CNRS, ONERA, Arts et Metiers Institute of Technology, Centrale Lille, UMR 9014 - LMFL - Laboratoire de Mécanique des Fluides de Lille - Kampé de Fériet, F-59000 Lille, France 
\aff{3} Department of Mechanical Engineering, Massachusetts Institute of Technology, 77 Massachusetts Avenue, Cambridge, 02139, MA, USA  
\aff{4} Institute of Fluid Mechanics (LSTM), Friedrich–Alexander–Universit{\"a}t (FAU) Erlangen--N{\"u}rnberg, DE-91058 Erlangen, Germany }

\begin{document}

\maketitle

\begin{abstract} 
Pipe flow controlled by streamwise-travelling waves of azimuthal wall velocity is studied using direct numerical simulations at a bulk Reynolds number $Re_b=4900$. A comprehensive analysis of drag reduction shows that the flow response differs fundamentally from that of channel flow. Under suitable forcing, pipe flow relaminarizes, whereas channel flow does not.
Depending on the control parameters, the flow exhibits the spatially localized turbulent state characteristic of transitional pipe flow, with turbulent puffs persisting at bulk Reynolds numbers up to three times higher than in the uncontrolled case. The bulk Reynolds number alone does not determine the onset of localization. Moreover, drag reduction, which alters the natural relation between bulk and friction velocities, is insufficient to identify a universal onset criterion.
An intermittency indicator based on the spatial variance of the cross-sectional turbulent kinetic energy relates the emergence of localized turbulence to the low wall friction produced by the control, although the correspondence is not one-to-one. Despite their qualitative resemblance to canonical turbulent puffs, the controlled puffs exhibit distinct properties; for example, their fronts may propagate faster than the bulk flow. 
Overall, this work provides a comprehensive characterization of the subcritical turbulent state and its turbulent puffs in controlled pipe flow, and offers a new perspective on control strategies that aim at flow relaminarization.
\end{abstract} 


\section{Introduction}
\label{sec:introduction}

Controlling turbulent flows for drag reduction remains an active area of research, driven by its economic and environmental relevance.
For parallel flows, where only friction drag is present, the open-loop control strategy based on streamwise-travelling waves of spanwise wall velocity (StTW) is known to be effective at reducing turbulent skin-friction drag and to produce large net savings \citep{quadrio-ricco-viotti-2009}. This method imposes a distribution of spanwise velocity at the wall, which varies with the streamwise position and time \cite[see ][for an in-depth review] {ricco-skote-leschziner-2021}. 
StTW have proved their drag reduction abilities at high $Re$ \citep{gatti-etal-2025} and in compressible flows \citep{gattere-etal-2024, ruby-foysi-2025}; for non-canonical flows, StTW have been applied to a transonic wing, yielding large improvements in the aerodynamic efficiency \citep{berizzi-etal-2025a}. 
StTW have also been found to improve the stability characteristics of the channel flow, delaying transition \citep{massaro-etal-2023}

StTW are primarily studied in plane channel flow. Although pipes are common in applications, circular pipe flow remains far less studied. A comprehensive characterization of StTW performance in pipe flow is not yet available and is therefore one of the goals of the current work.
\cite{quadrio-sibilla-2000} studied uniform wall oscillation, which represents the limit case of a travelling wave with infinite wavelength, in a pipe and determined its potential for drag reduction. \cite{liu-etal-2022} considered a stationary wave by changing the control amplitude and wavelength. Recently, the drag-reducing potential of purely rotating pipe  (\textit{i.e.} wave with infinite wavelength and oscillation period) was assessed by \cite{xiao-etal-2024}. 
The first experimental investigation of StTW was carried out in the pipe geometry by \cite{auteri-etal-2010} at the same $Re$ as the original numerical study by \cite{quadrio-ricco-viotti-2009}, namely $Re_b=4900$, where $Re_b = U_b D / \nu$ with $D$ the pipe diameter, $U_b$ the bulk velocity, and $\nu$ the kinematic viscosity. Their actuation device, based on independently rotating slabs, implemented a spatially discrete analogue of travelling waves, and two forcing wavelengths were examined.
Recently, \cite{gallorini-quadrio-2024} performed direct numerical simulations (DNS) of the same pipe flow controlled with discrete StTW. To enable a direct comparison, they kept the same relatively low $Re_b$ of the experiment, and discussed how differences between experiments and simulations (previously available for the planar geometry only) are caused by the spatially discrete waveform of the azimuthal forcing. 
An important additional difference was identified: depending on the control parameters, the flow in the simulation may either fully relaminarize or develop spatially localized turbulence. 

Hence, beyond the obvious interest in drag-reduction data for the relatively unexplored pipe geometry, this study examines drag-reduced flows at low Reynolds numbers, offering a unique perspective on the transition to turbulence in a controlled pipe flow.
The occurrence of localized turbulence at $Re_b=4900$ is, in fact, unexpected, as this value far exceeds the value of the critical Reynolds number $Re_c=2040$ reported by \cite{avila-etal-2011} for the onset of localized turbulence in canonical pipe flow.
Pipe flow transition has been studied intensively since the pioneering experiments of \cite{Reynolds-1883}, yet some aspects remain a matter of debate. 
In canonical pipe flow, transition is subcritical: the Hagen--Poiseuille parabolic velocity profile is stable to infinitesimal perturbations for any value of $Re$ \citep{schmid-henningson-2001}. 
Numerical simulations have indeed shown that laminar pipe flow disturbed by infinitesimal perturbations remains stable up to $Re = 10^7$ \citep{meseguer-trefethen-2003}.
Experiments, however, have not reported laminar flow at such high Reynolds numbers \citep{vandoorne-westerweel-2008}, indicating that pipe flow transition is triggered by finite-amplitude disturbances. During transition to turbulence, these localized structures proliferate: their average growth rate increases with $Re$, first exceeding their decay rate, so that puffs are surrounded by laminar flow \citep{avila-etal-2011}, and eventually, at higher $Re$, conventional space-filling turbulence is obtained \citep{moxey-barkley-2010, avila-hof-2013}.

The Reynolds number $Re_b$ determines the boundary between growing and decaying puffs and marks the onset of turbulence. Isolated puffs have been observed at $Re_b \approx 1500$ \citep{hof-etal-2005}. 
At higher $Re_b$, the turbulent clusters expand at a rate that, after the initial formation time, is independent of $Re_b$ \citep{barkley-etal-2015}; these are called slugs \citep{lindgren-1957, wygnanski-champagne-1973}. 
Puffs have coherent features, travel at a speed slightly smaller than $U_b$, and have a well-defined streamwise length, independent of the details of the disturbance that creates them. 
They emerge sharply from the quiet, laminar state at their upstream side, whereas the laminar state downstream recovers slowly. 
Over recent decades, turbulent puffs have been characterized in detail both experimentally and numerically \cite[see][and references herein]{avila-barkley-hof-2023}. 
\cite{wygnanski-sokolov-friedman-1975} experimentally studied pipe flow transition in the range $2000 \leq Re_b \leq 2700$, and observed non-decaying puffs, regardless of the initial perturbation, and therefore self-sustained turbulence at $Re_b \geq 2200$. 
Whether turbulent puffs cease to decay at a well-defined critical Reynolds number or remain inherently transient at low $Re_b$ is an ongoing debate.
 Experiments \citep{peixinho-mullin-2006} and numerical simulations \citep{willis-kerswell-2007} have found that turbulent puffs relaminarize at low $Re_b$, showing power-law scaling of the transient decay half-life and suggesting a boundary crisis. 
At the same time, \cite{hof-etal-2006} suggested a different scenario, noting that the characteristic time of a turbulent puff does not diverge; instead, it increases exponentially with the Reynolds number, as numerically pointed out by \cite{faisst-eckhardt-2004} up to $Re_b=2100$. This finding was later confirmed both experimentally \citep{hof-etal-2008, delozar-hof-2009} and numerically \citep{avila-willis-hof-2010}.
Determining the critical value $Re_c$ of the Reynolds number at which turbulence becomes sustained has been a long-standing challenge. Initially, $Re_c$ was defined as the point where the lifetime of a turbulent cluster diverges. 
Later, proliferation/splitting processes were identified as the key mechanism leading to sustained turbulence \citep{moxey-barkley-2010} and, by comparing the characteristic timescales of puff decay and splitting, \cite{avila-etal-2011} identified a critical Reynolds number $Re_c=2040$, at which the decay and splitting timescales are equal.
This finding was later confirmed by \cite{mukund-hof-2018}, who demonstrated that puffs relaminarize for $Re_b < 2020$, while turbulence becomes sustained for $Re_b > 2060$. 

In the studies discussed above, the onset of transition in pipe flow is universally characterized by the bulk Reynolds number $Re_b$. In canonical pipe flow, a given value of $Re_b$ corresponds to a unique K\'arm\'an number $R^+$, defined as $R^+ = R u_\tau / \nu$, where $R$ is the pipe radius, and $u_\tau$ is the friction velocity, i.e., the square root of the wall friction over the fluid density.
$R^+$ is analogous to the friction Reynolds number $Re_\tau$, which is widely used to describe plane channel flow.
However, the link between $Re_b$ and $Re_\tau$ is altered when skin-friction drag reduction  lowers $Re_\tau$ for a given $Re_b$, or, conversely, increases $Re_b$ for a given $Re_\tau$ \citep{hasegawa-quadrio-frohnapfel-2014}. 
Therefore, the present study systematically examines the transitional and turbulent states in pipe flow with StTW, where the usual relationship between $Re_b$ and $R^+$ is broken. DNS is used to explore a broad region of the control parameter space, with the Reynolds number fixed at $Re_b = 4900$, well above the value required for the uncontrolled flow to produce conventional, space-filling turbulence.

The paper is organized as follows. §\ref{sec:sims} describes the numerical methods and computational procedures used to generate the DNS dataset. An overview of the drag reduction results is presented in §\ref{sec:results-dr}, followed by a discussion of the localized turbulence state in §\ref{sec:results-localized}. Finally, §\ref{sec:conclusion} presents  concluding remarks.
 \section{Simulations}
\label{sec:sims}

Direct numerical simulations of the flow in a cylindrical pipe are carried out by numerically integrating the incompressible Navier--Stokes equations, formulated in primitive variables and cylindrical coordinates. 

The DNS solver is written in the CPL programming language \citep{luchini-2020, luchini-2021} and is derived from the Cartesian solver originally introduced by \cite{luchini-quadrio-2006}; additional details are provided in \cite{gallorini-quadrio-2024}.
The equations are written in terms of velocity $\vect{u} = (u_x,u_r,u_\theta)$ and pressure $p$, where $u_x$, $u_r$, and $u_\theta$ denote the velocity components in the axial, radial, and azimuthal directions, respectively. 
The periodic azimuthal and axial directions are spectrally discretized, while compact second-order finite differences are used in the radial direction. No-penetration and no-slip conditions are applied at the pipe wall. A non-homogeneous Dirichlet condition on the azimuthal velocity component is applied to prescribe the control action. Regularity conditions are imposed at the pipe axis \citep{lewis-bellan-1990}.
The time advancement scheme employs a partially implicit approach. The viscous term is discretised using the implicit Crank--Nicolson scheme, while the convective term is treated using the three-substep, low-storage Runge--Kutta scheme described by \cite{rai-moin-1991}. 
The number of azimuthal Fourier modes is constrained by the resolution near the pipe wall. Since this would lead to unnecessarily high resolution toward the pipe axis and to a consequent severe time step limitation, the code smoothly reduces the number of azimuthal modes as the pipe axis is approached.

The simulations are carried out at a constant flow rate, adjusting the axial pressure gradient at every time step, as described by \cite{quadrio-frohnapfel-hasegawa-2016}. Therefore, the bulk Reynolds number $Re_b=U_b D /\nu=4900$ is constant during the simulations. 
For the reference (uncontrolled) turbulent pipe flow, $Re_b=4900$ corresponds to $R^+=170$. 
Since the axial domain length $L_x$ is one of the most critical discretization parameters, a dedicated analysis is conducted in Appendix \ref{sec:domain}, varying $L_x$ from $8.5D$ to $122.5D$. The results indicate that $L_x \simeq 30D$ is sufficient to accurately reproduce the quantities of interest, and $L_x=30D$ is therefore employed for the baseline set of cases. Note that $L_x=30D$ exceeds the lengths typically used for turbulent pipe flow simulations \citep{kim-adrian-1999, wu-moin-2008}. 
The axial and azimuthal directions are discretized using $1364 \times 192$ Fourier modes. After dealiasing, the grid spacing is $\Delta x^+ = 5$ and $r \Delta \theta^+ = 3.7$ at the pipe wall for the uncontrolled case, where $(\cdot)^+$ indicates quantities expressed in viscous units. Note that drag reduction implies finer spatial resolution in actual viscous units.
The number of azimuthal modes $N_\theta$ varies with the radial direction: it remains constant at $N_\theta=192$ from the wall ($r=D/2$) to $r = 0.4R$, then it decreases linearly, as described by \cite{gallorini-quadrio-2024}, to reach $N_\theta=4$ at the pipe centerline. 
The radial grid spacing increases smoothly from the wall toward the pipe axis, following a hyperbolic tangent distribution: it is $\Delta r^+ = 0.7$ at the wall, and $\Delta r^+ = 2.4$ at the centerline in the reference flow. 
The timestep $\Delta t$ is dynamically adjusted to ensure that the Courant--Friedrichs--Lewy (CFL) number equals unity, with an average value of $\Delta t^+ = 0.1$. Additional simulations were conducted with the smaller value  CFL$=0.5$, to confirm the absence of any noticeable difference in the results.

A large dataset is built, with more than one thousand (1184) simulations of pipe flow controlled by StTW. 
For all cases, the initial condition is an unforced turbulent flow in statistical equilibrium. The applied wall forcing is
\begin{equation}
  u_\theta(x,t) = A \sin \left( k_x x - \omega t \right),
  \label{eq:control}
\end{equation}
containing three parameters: the forcing amplitude $A$, the wavenumber $k_x$, and the angular frequency $\omega$.
In this study, two of the three parameters are varied, with the amplitude fixed at $A/U_b=1$ (or $A^+=14.5$), the same value used by \cite{auteri-etal-2010} and \cite{gallorini-quadrio-2024}.
The baseline pipe length $L_x=30 D$ is slightly adjusted to always accommodate an integer number of waves. The simulations are run for $500 D/U_b$. Statistics are computed after discarding the first half of the time history to remove the initial transient, during which the flow adjusts to the controlled state.
For the analyses described in \S~\ref{sec:puffAnalysis} and Appendix \ref{sec:domainL}, results are obtained by averaging over instantaneous velocity fields saved over a time window of $tU_b/D = 250$ after eliminating the transient.

Results are made dimensionless either in external units (with $U_b$ and $D$ as the velocity and length scales) or in inner units (with $u_\tau$ and $\nu/u_\tau$). Since $U_b$ is fixed, drag reduction changes the value of $u_\tau$. The superscript $(\cdot)^+$ is used for quantities expressed in friction units of the uncontrolled case, whereas the superscript $(\cdot)^*$ indicates scaling based on the actual friction velocity of the controlled case.

Several averaging operators are employed. $\aver{u}(r)$ indicates the average of the field $u(x,r,\theta,t)$ over time and the two homogeneous directions
\begin{equation}
\aver{u}(r) = \frac{1}{2\pi L_xT}\int_0^T\int_0^{L_x}\int_0^{2\pi} u(x,r,\theta,t) \, \text{d}\theta \text{d}x \text{d}t,
\end{equation}
where the time is measured after discarding the transient. Similarly, we define the azimuthal average at time $t$ as:
\begin{equation}
\thaver{u}(x,r,t) = \frac{1}{2\pi}\int_0^{2\pi} u(x,r,\theta,t) \, \text{d}\theta .
\end{equation}

The notation $\paver{u}(r)$ indicates the average of $u(x,r,\theta,t)$ inside a puff, i.e., within an instantaneous turbulent structure as defined in \S~\ref{sec:results-localized}. 
If a structure is defined with a local streamwise axis $x'$ such that its extension ranges from $x'=0$ to $x'=\ell(t)$, the puff-averaged axial velocity is:
\begin{equation}
\paver{u}(r) = \frac{1}{2\pi \ell T}\int_0^T\int_0^\ell \int_0^{2\pi} u(x',r,\theta,t) \, \text{d}\theta \text{d}x' \text{d} t .
\end{equation} 

The availability of a puff-averaged friction velocity implies that, besides the $*$ scaling built with the actual-global friction velocity, an additional actual-local scaling can be used, which will be indicated with a double asterisk superscript, namely $(\cdot)^{**}$.
 \section{Drag reduction}
\label{sec:results-dr}

This section presents the drag reduction performance of StTW in pipe flow, which is still missing as outlined in \S\ref{sec:introduction}. 
The current dataset is designed to cover very low streamwise wavenumbers, since information on this region of parameter space is quite limited for the plane channel case, owing to the high computational cost. The dataset includes cases in which $k_x$ corresponds to a forcing wavelength of up to $27D$. For comparison, channel flow studies typically use domains with streamwise lengths of about $L_x \approx 6D$ (interpreting $D$ as the gap between the two walls).

\begin{figure}
\includegraphics[width=\textwidth]{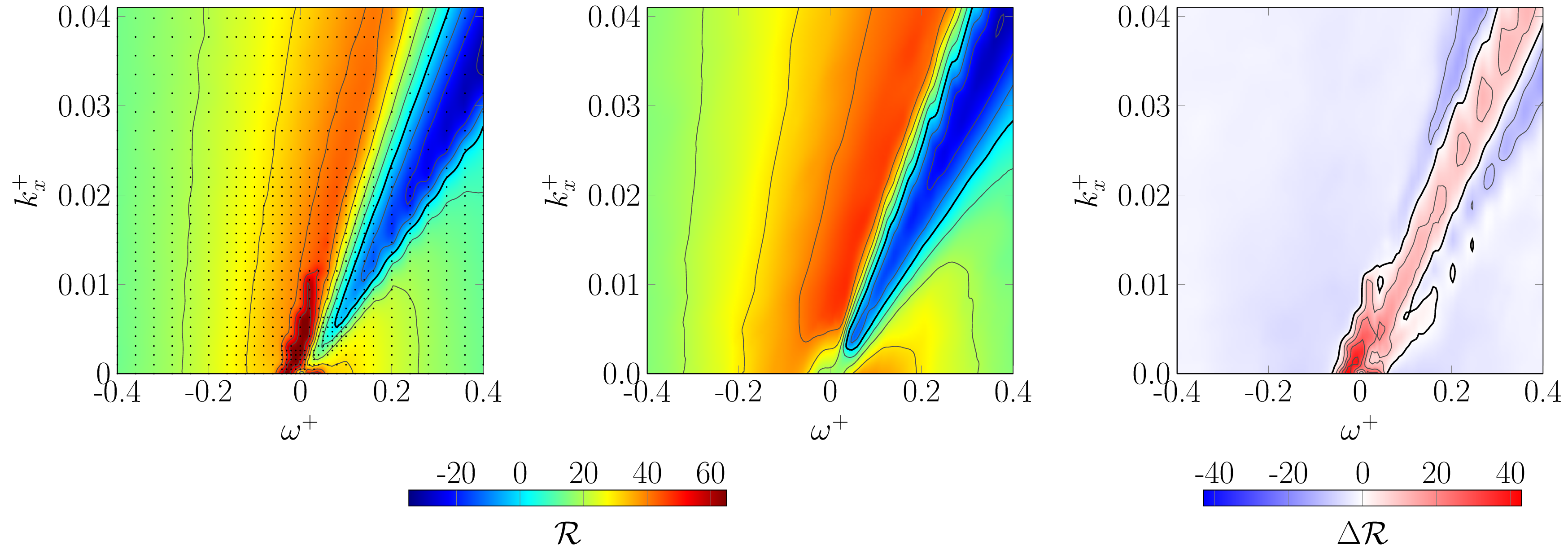}
\caption{Maps of the percentage drag reduction rate $\mathcal{R}$ in the $\omega-k_x$ plane for $A^+ = 14.5$. Present results (left) are compared with plane channel flow data (center), extrapolated from $Re_\tau=200$ to $Re_\tau=170$. Contours are spaced by $10\%$ intervals; the thick line indicates the zero contour level. The right panel shows the difference $\Delta \mathcal{R}$ between pipe and channel. In the left panel, the dots show where simulations have been carried out.}
\label{fig:DRMap}
\end{figure}

The drag reduction rate $\mathcal{R}$ is quantified by the percentage change in the friction coefficient $C_f$ as:
\begin{equation}
  \mathcal{R} =  100 \left( 1 - \frac{C_f}{C_{f,0}} \right) ,
\end{equation}
where the subscript $(\cdot)_0$ refers to quantities computed for the reference case. The coefficient $C_f$ is defined as:
\begin{equation}
  C_f = \frac{2 \tau_w}{\rho U_b^2},
\end{equation}
where $\tau_w$ is the mean streamwise shear stress at the pipe wall:
\begin{equation}
  \tau_w = - \mu \frac{\partial \aver{u_x}}{\partial r} \Bigl|_{r=D/2}
\end{equation}
and $\mu$ is the dynamic viscosity. 
Figure \ref{fig:DRMap} provides an overview of $\mathcal{R}$ in the $(\omega^+, k_x^+)$ plane, in comparative form between channel and pipe.
The dots on the map in the left panel indicate the frequency-wavenumber combinations at which the pipe flow simulations are carried out. Regions where drag reduction rates exhibit strong gradients, particularly near the transition between drag reduction and drag increase at low $k_x^+$, are locally refined by adding sampling points. 
Note that the point at $\omega^+=0$ and $k_x^+=0$ represents a wave with infinite period and wavelength, therefore a uniformly rotating pipe, which in principle produces drag reduction \citep{xiao-etal-2024}. Per the definition of equation \ref{eq:control}, the rotation velocity is $u_\theta(x,t)=A\sin(0)=0$ in our case, corresponding to a motionless pipe, hence $\mathcal{R}=0$. The cylindrical case differs from the planar one, where a uniform wall velocity may result, depending on the motion law for the wall, in the two walls moving in the same or in opposite directions.
For comparison, the central panel shows $\mathcal{R}$ for a plane channel at $A^+=14.5$, obtained by extrapolating the dataset of \cite{gatti-quadrio-2016} at $Re_\tau=200$. The effect of the spanwise forcing can be converted into a Reynolds-independent measure, i.e., the shift $\Delta B^+$ of the logarithmic layer of the mean velocity profile
\begin{equation}
\Delta B^+ = \sqrt{\frac{2}{C_{f,0}}}[(1-\mathcal{R}/100)^{-1/2}-1]-\frac{1}{2\kappa}\ln{(1-\mathcal{R}/100)},
\label{eq:DBp}
\end{equation}
where $\kappa$ is the von K\'arm\'an constant. By solving equation \ref{eq:DBp} and knowing $\Delta B^+$, the value of $\mathcal{R}$ can be derived. The friction coefficient $C_{f,0}$ of the uncontrolled flow is the only term from equation \ref{eq:DBp} that depends on the Reynolds number. Its value has been obtained with a channel flow simulation at $Re_\tau = 170$, corresponding to $Re_b \approx 5300$, while a value of $\kappa=0.41$ has been used for the von K\'arm\'an constant.

The map is, as expected, qualitatively similar to that of plane channel flow at the same Reynolds number. 
A quantitative comparison reveals higher drag reduction in the pipe flow, where a maximum of $\mathcal{R} \approx 65\%$ is reached, compared to $\approx 51\%$ in the channel.
Note that $\mathcal{R} \approx 65\%$ represents the theoretical maximum at this Reynolds number, achieved only after complete relaminarization. When laminarization occurs, drag reduction cannot be interpreted in terms of a shift of the logarithmic layer of the mean velocity profile.
Existing literature confirms that turbulent pipe flow tends to relaminarize more readily than channel flow \citep{liu-etal-2022}, a difference that may be related to different stability properties of the base flow controlled with different control parameters.
A second observation is that the region of largest $\mathcal{R}$ in the pipe shifts toward lower wavenumbers, extending down to the $k_x=0$ axis where the forcing is purely temporal (oscillating pipe). While $\mathcal{R}$ is symmetric along the horizontal axis, for very small $k_x$ the left quadrant (backward-travelling waves) exhibits higher drag reduction, as in the plane channel.
The drag-increasing region has a triangular shape in both cases. 
For the pipe, the vertex of the triangle defined by the zero contour level is at $\omega^+ \approx 0.075$ and $k_x^+ \approx 0.005$, which is slightly shifted away from the origin than in the plane channel. This position is consistent with the larger $\mathcal{R}$ in the pipe at low wavenumbers. The maximum drag increase observed across the examined range of parameters is $35\%$ in the pipe, compared to $30\%$ in the plane channel.
The drag-increase triangular region is centered on a phase speed of approximately $10 u_\tau$, in agreement with the original observation by \cite{quadrio-ricco-viotti-2009} that linked this phase speed to the average convection velocity of near-wall turbulent fluctuations and the turbulence structures sustaining them.
The rightmost panel of figure \ref{fig:DRMap} plots the difference $\Delta \mathcal{R}$ in drag reduction rates between pipe and channel. Since the channel-flow data result from an extrapolation procedure, we expect the $\Delta \mathcal{R}$ map to capture qualitative differences and highlight the regions where differences are most significant, rather than providing their exact quantification. Despite the overall similarity, local differences can be substantial, reaching over $40\%$. These large differences are concentrated near the origin of the plane, where laminarization occurs in the pipe flow.

Interestingly, outside of the high-$\mathcal{R}$ region, the channel flow shows higher drag reduction and lower drag increase than the pipe. 
To rule out any unintended effect of extrapolating the plane-geometry data in the dataset of \cite{gatti-quadrio-2016}, we have carried out an additional simulation for the plane channel flow at exactly the same $Re_\tau=170$. For the parameter values $k_x^+=0.02$, $A^+=14.5$ and $\omega^+=-0.2$, the resulting $\mathcal{R}=24\%$ is consistent with the value from equation \ref{eq:DBp} ($\mathcal{R}=23\%$), and, compared to $\mathcal{R}=22\%$ in the controlled pipe flow, confirms the slightly larger effectiveness of StTW at high frequencies in the planar geometry. Given the vicinity of the two values of $\mathcal{R}$, we computed the statistical uncertainty $\delta R$ with the procedure described in Appendix \ref{sec:domainL}, finding a value of $\delta R = 0.6\%$ for the channel and $\delta R = 0.4\%$ for the pipe, which indicates that the difference $\Delta \mathcal{R}$ is statistically significant.
 
\begin{figure}
\includegraphics[width=\textwidth]{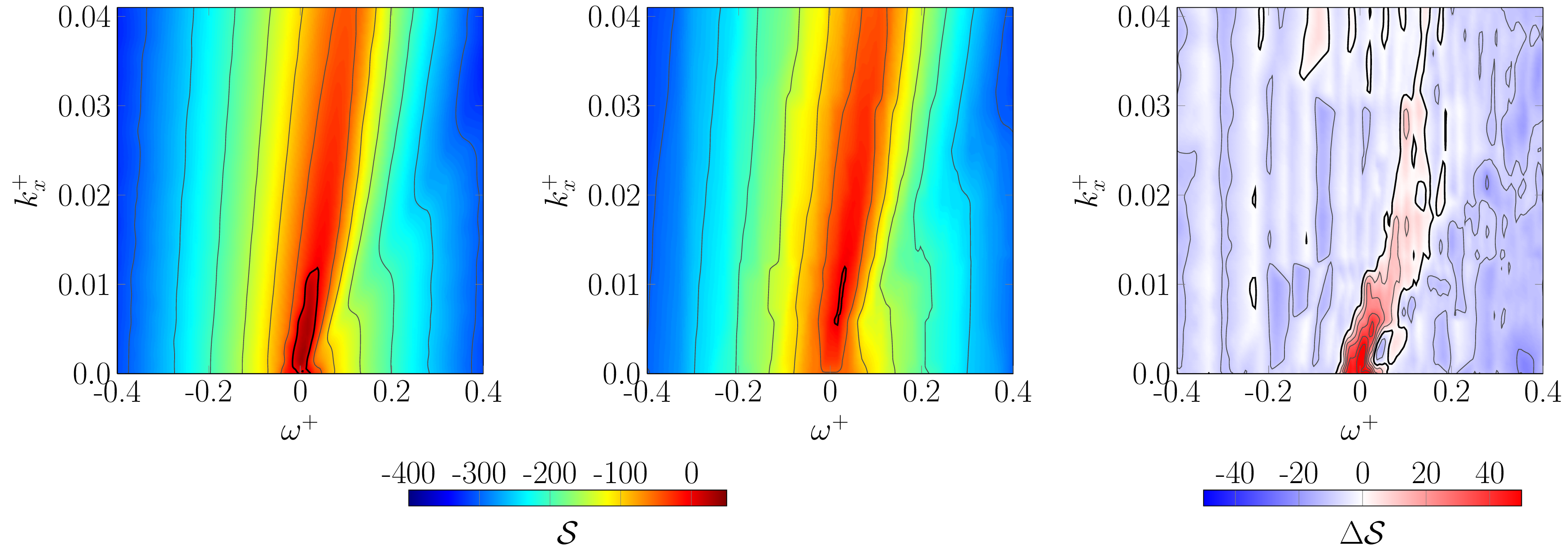}
\caption{Maps of the net power saving $\mathcal{S}$ in the $\omega-k_x$ plane for $A^+ = 14.5$. Present results (left) are compared with plane channel flow data (center), extrapolated from $Re_\tau=200$ to $Re_\tau=170$. Contours are spaced by unit intervals; the thick line indicates the zero level. The right panel shows the difference $\Delta \mathcal{S}$ between pipe and channel.}
\label{fig:SMap}
\end{figure}

Since StTW is an active technique and requires power to operate, the comparison between pipe and channel flows should be extended to consider the net power budget. The net saving $\mathcal{S}$ is computed by accounting for the control power $\mathcal{P}_{in}$, i.e. the power cost required to obtain $\mathcal{R}$
\begin{equation}
\mathcal{S} = \mathcal{R}-\mathcal{P}_{in},
\end{equation}
where $\mathcal{P}_{in}$ is the input power that an ideal control system transfers to the flow, normalized by the pumping power $P_0$: 
\begin{equation}
\mathcal{P}_{in} = \frac{\mu}{P_0} \aver{ \left[ u_\theta \left( \frac{\partial u_\theta}{\partial r} - \frac{u_\theta}{r} \right) \right]_{r=D/2} }.
\end{equation}

Figure \ref{fig:SMap} compares the map of $\mathcal{S}$ for pipe and for plane channel flow. Channel data are computed, following \cite{ricco-quadrio-2008}, under the assumption that the change in friction only affects $U_b^+$. The power spent then becomes $\mathcal{P}_{in} \propto (U_b^+)^2$, and empirical correlations yield $U_b^+ = 0.5 Re_b / Re_\tau = 7.715Re_\tau^{0.136}$.
Note that the control amplitude used here is relatively large and differs from the amplitude that yields maximum net savings for the channel flow, which is approximately $A^+ \approx 6$ \citep{gatti-quadrio-2016}.
Therefore, a positive net saving is nearly absent (4\%) in the plane channel case, while for pipe flow the area with $\mathcal{S} > 0$ is wider and the maximum $\mathcal{S}$ is higher (28\%) because of laminarization. 
The reliability of these findings is reinforced by the additional simulation for the plane channel flow at $k_x^+ = 0.02$ and $\omega^+ = -0.2$, which gives a net saving of $\mathcal{S} = -200\%$, compared to $-220\%$ for the pipe flow, in agreement with the trend of figure \ref{fig:SMap}. 
We stress again, however, that the channel flow data derive from extrapolation, and therefore such comparisons should be interpreted qualitatively. For example, the $P_{in}$ obtained for the channel flow DNS at $k_x^+ = 0.02$ and $\omega^+ = -0.2$ differs from the extrapolated value by $4\%$.
We have also confirmed that the trends discussed here for both $\Delta \mathcal{R}$ and $\Delta \mathcal{S}$ remain valid when actual viscous scaling is used instead of the viscous scaling based on the friction velocity of the uncontrolled case.

\begin{figure}
\centering
\includegraphics[width=0.75\textwidth]{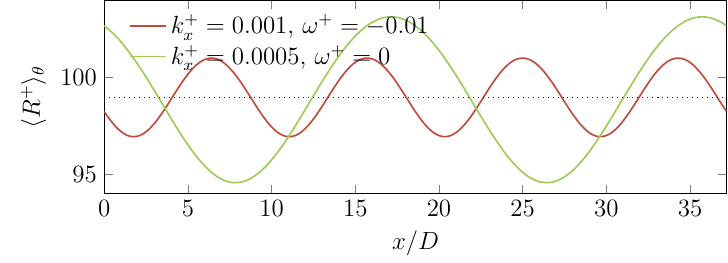}
\caption{Local wall friction $\langle R^+\rangle_\theta$ for different control parameters resulting in a full laminarization of the flow. The dotted line indicates the friction level for the unforced laminar flow.}
\label{fig:pipe_lam} 
\end{figure}

Relaminarization by StTW in pipe flow is further discussed by presenting the local wall friction in figure~\ref{fig:pipe_lam} for two relaminarized cases. The figure shows the instantaneous local value of the K\'arm\'an number $\thaver{R^+}(x)=\langle{\frac{D}{2}(-\frac{1}{\nu}\frac{\partial u}{\partial r}|_{r=D/2})^{1/2}\rangle}_\theta$.
An evident streamwise modulation of the flow is observed neither in the plane counterpart nor in the purely oscillating pipe \citep{coxe-peet-adrian-2022}. 
For a laminar flow in a plane channel, it was shown by \cite{quadrio-ricco-2011} that the spanwise motion of the generalized Stokes layer decouples from the streamwise one, and therefore the wall slope of the streamwise velocity profile does not vary with $x$.
For laminar flow in a pipe, however, the Navier--Stokes equations written in cylindrical coordinates show that the viscous terms become coupled. Therefore, under StTW with $k_x \ne 0$, the laminar solution depends on time, wall-normal (radial) and streamwise coordinates, while remaining azimuthally homogeneous: the azimuthal velocity component couples with the radial and streamwise ones.
Such a streamwise-varying solution resembles the one observed, both in the channel \citep{quadrio-ricco-viotti-2009} and the pipe \citep{gallorini-quadrio-2024}, when StTW  produce drag increase.
 \section{Localized turbulence}
\label{sec:results-localized}

At $Re_b=4900$, the controlled pipe flow generally resembles a conventional turbulent wall flow with modified (increased or decreased) friction. However, for certain control parameters, the flow becomes fully laminar and, more interestingly, sometimes exhibits spatially localized turbulence. In this Section, we focus on the  localized turbulence state observed in a controlled pipe flow.

\subsection{Identification}
\label{sec:identification}

A criterion is needed to discriminate between space-filling turbulence and transitional, localized turbulence with puffs. 
In canonical pipe flow, several approaches exist to detect turbulent puffs. 
The instantaneous cross-plane turbulent kinetic energy $q(x,t)$, employed for instance by \cite{moxey-barkley-2010}, is commonly used to detect puffs. The quantity $q(x,t)$ is defined as the sum of squares of the radial and azimuthal velocity fluctuations (indicated by the superscript $'$), over the cross-sectional plane with area $A$. This averaging operator is denoted by $\aver{\cdot}_A$:
\begin{equation}
  q(x,t) = \frac{4}{\pi D^2} \int_0^{D/2} \int_0^{2 \pi} \left( {u_r'}^2+{u_\theta'}^2 \right) \ r \text{d}r \text{d}\theta = \aver{ {u_r'}^2+{u_\theta'}^2 }_A .
\end{equation}
The flow exhibits streamwise modulation due to the forcing in the azimuthal component. To mitigate it, the forced harmonic is removed from the fluctuating field.

\begin{figure}
\includegraphics[width=\textwidth]{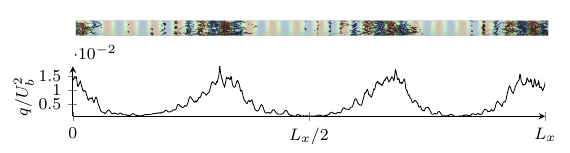}
\caption{Instantaneous visualization of localized turbulence along the pipe. The control parameters $k^+=0.0126$ and $\omega^+=0.08$ (case I of table \ref{tab:puffChar}) correspond to a large drag reduction of $\mathcal{R}=31\%$. Top: instantaneous configuration of the control (represented by the background colour, which indicates the azimuthal velocity of the wall) along with turbulence structures, visualized as isosurfaces of $\lambda_2^+=-0.06$. Bottom: evolution of $q$ along the axial direction.}
\label{fig:locTurb}
\end{figure}

The streamwise evolution of $q$ for parameters $k^+=0.0126$ and $\omega^+=0.08$ (case I of table \ref{tab:puffChar}, with a large drag reduction of $\mathcal{R}=31\%$) is shown in figure \ref{fig:locTurb}, along with a visualization of the corresponding instantaneous turbulent structures.
Using the $\lambda_2$ criterion \citep{jeong-hussain-1995}, these vortical structures appear as three localized clusters, separated by quiescent regions where turbulent fluctuations are significantly weaker. 
Correspondingly, $q(x)$ peaks at the center of the clusters and gradually decreases to nearly zero in the gaps between adjacent clusters. 
In quiescent regions, minor oscillations persist due to imperfect forcing removal; nonlinearity transfers energy to higher harmonics.

To quantify localization across all cases, $q(x,t)$ must be compressed into a single metric.
Several alternatives have been tested, finding that the variability of $q$ along the streamwise direction is best quantified by the indicator function $\mathcal{I}$, defined as the root-mean-square value $\sigma_q$ of the space--time fluctuations of $q$ over the entire time history, normalized by the bulk velocity squared:

\begin{equation}
  \mathcal{I} = \frac{\sigma_q}{U_b^2} .
\label{eq:I}
\end{equation}

\begin{figure}
\centering
\includegraphics[width=0.8\textwidth]{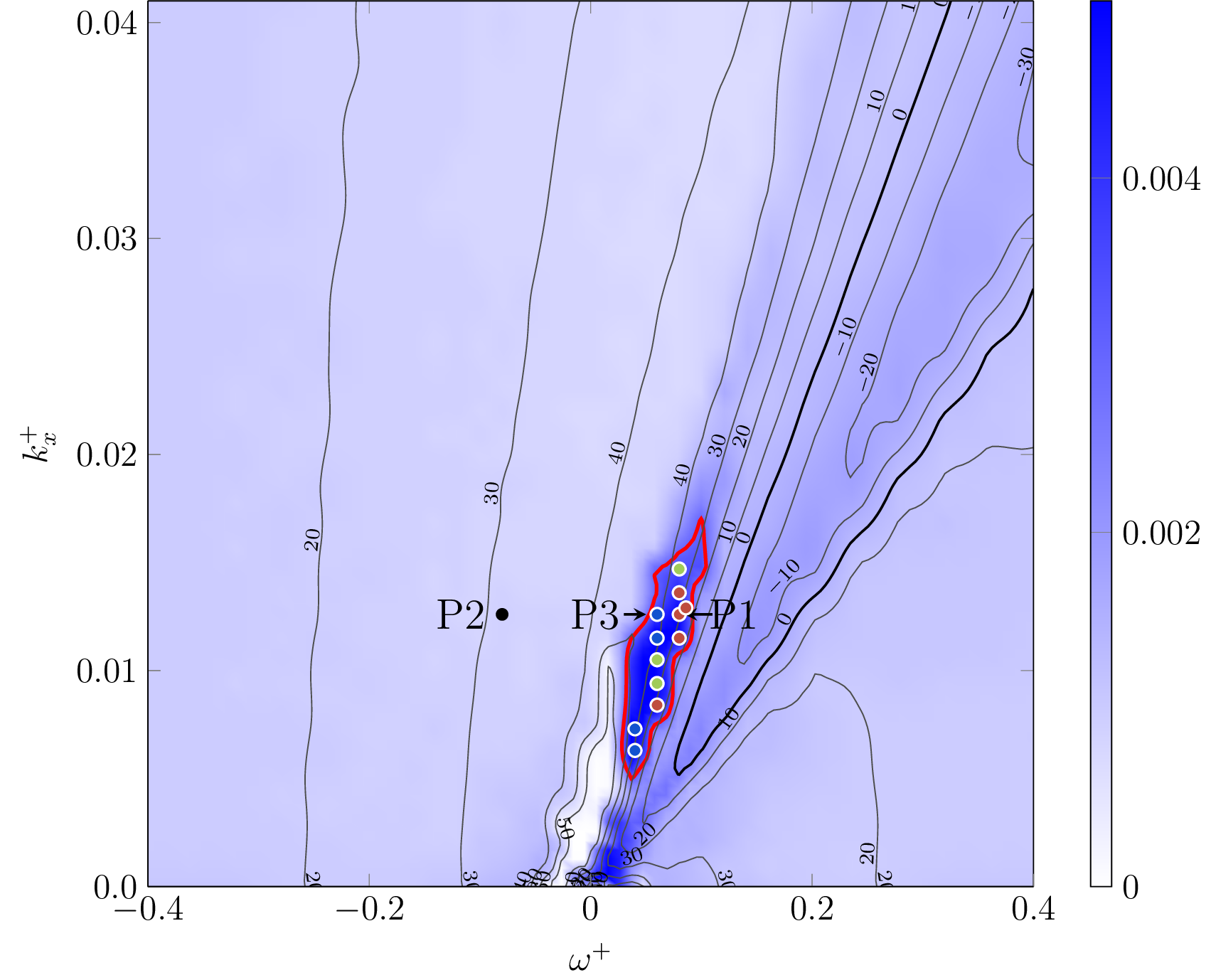}
\caption{Map of the intermittency indicator $\mathcal{I}$ in the $\omega-k_x$ plane for $A^+ = 14.5$. Contours of $\mathcal{R}$, spaced by $10\%$ intervals, are overlaid to visualize the relation between $\mathcal{I}$ and drag reduction. The color scale is saturated above $\mathcal{I}=0.005$. The colored dots indicate where localized turbulence is observed, with the number $n$ of turbulent regions encoded by color: $n=1$ blue, $n=2$ green, and $n=3$ red.
}
\label{fig:locTurbMap}
\end{figure}

High values of $\mathcal{I}$ indicate spatially localized turbulence: when the flow contains alternating laminar and turbulent regions, $q(x)$ ranges from values close to zero in quasi-laminar zones to large values in the turbulent core of the puffs. 
Figure \ref{fig:locTurbMap} shows a colormap of $\mathcal{I}$ as a function of the control parameters, with drag reduction contours overlaid.
High values of $\mathcal{I}$ occur at relatively low wavenumbers ($k_x^+<0.02$) and positive frequencies ($0 < \omega^+ < 0.1$): spatially localized turbulence occurs mostly for forward-travelling waves.
As expected, large drag reduction correlates with localized turbulence.
However, the opposite is not true: large drag reduction does not necessarily imply localized turbulence. For instance, in much of the negative-frequency region, turbulence remains space-filling while drag reduction is large.
Moreover, maximum $\mathcal{R}$ (corresponding to relaminarization) yields very low $\mathcal{I}$ values.

We now examine three representative cases: P1, P2, and P3 in figure \ref{fig:locTurbMap}. 
Points P1 (at $\omega^+=0.08$, $k_x^+=0.0126$, I in table \ref{tab:puffChar}) and P2 (at $\omega^+=-0.08$, $k_x^+=0.0126$) produce the same drag reduction $\mathcal{R}=31\%$. Their wavenumber is identical, and their frequencies are opposite, yet their flow states differ: localized turbulence for P1 (downstream-travelling wave) and space-filling turbulence for P2 (upstream-travelling). The third point P3 (at $\omega^+=0.06, k_x^+=0.0126$, H in table \ref{tab:puffChar}) has the same wavenumber and flow state (localized turbulence) as P1, but achieves a larger drag reduction, with $\mathcal{R}=43\%$.

\begin{figure}
\centering
\includegraphics[width=0.75\textwidth]{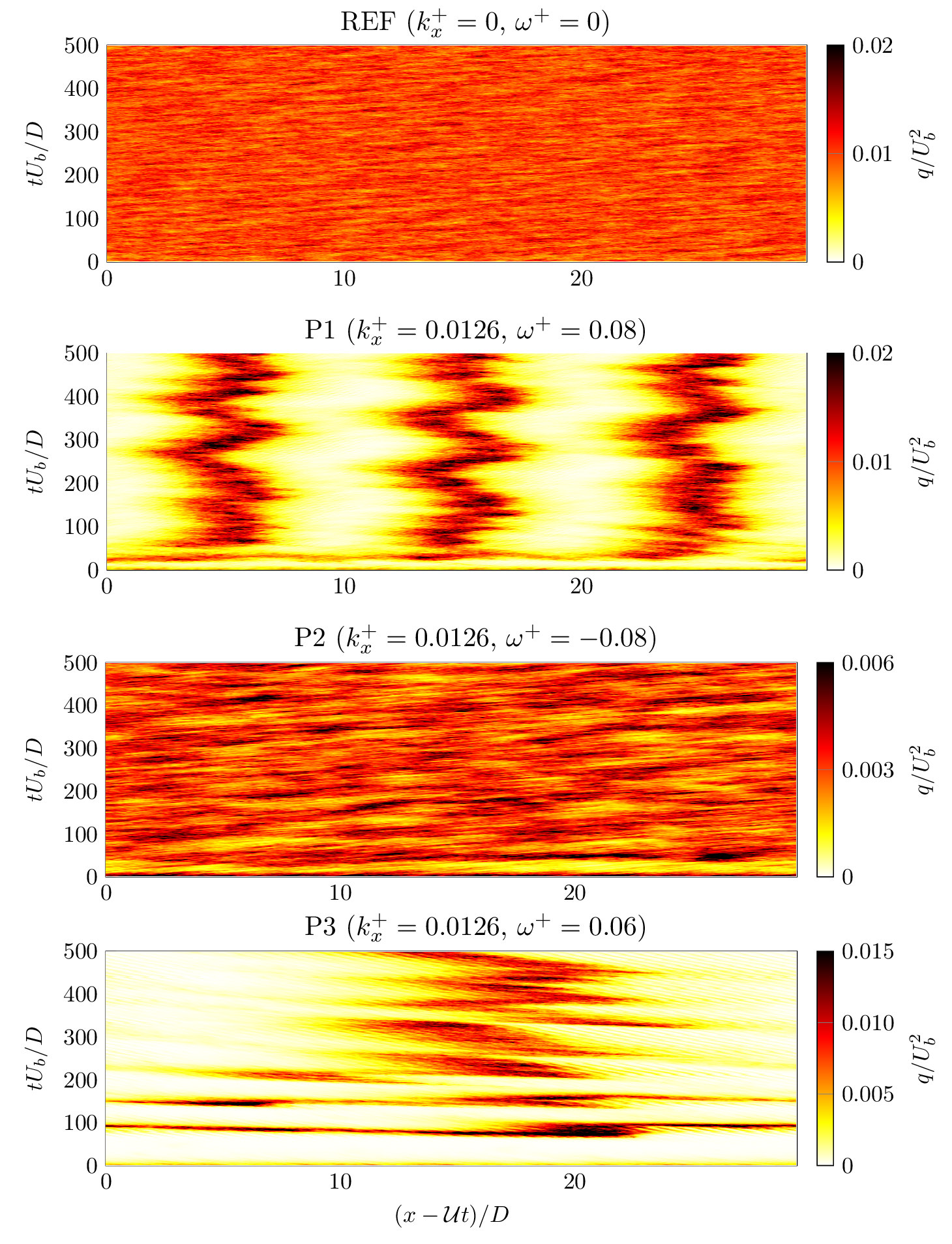}
\caption{Space--time evolution of $q(x,t)$. From top to bottom: REF, P1, P2, and P3. Data are represented in the  convecting reference frame at a speed $\mathcal{U}$: $\mathcal{U}=U_b$ for REF anf P2, $\mathcal{U}=1.12U_b$ for P1, and $\mathcal{U}=1.4U_b$ for P3. }
\label{fig:qxt}
\end{figure}

Figure \ref{fig:qxt} is a spatio-temporal plot of the function $q(x,t)$ for the three cases, all starting from the same uncontrolled initial condition. The same quantity for the flow in the absence of a control, denoted as REF, is also reported for comparison. Such space-time representation has been often used to examine the characteristics of transitional turbulent puffs in the canonical pipe flow \citep{moxey-barkley-2010, avila-etal-2011}. The puffs are observed in a convecting reference frame moving at speed $\mathcal{U}$. For P1 and P3, presenting localized turbulence, $\mathcal{U}$ is the mean convective velocity of the turbulent region; for P2 and REF, $\mathcal{U}=U_b$ as in \cite{moxey-barkley-2010}.

After control is applied at $t=0$, $q(x,t)$ undergoes a characteristic transient phase, already described by \cite{ricco-etal-2012} for the friction level, where fluctuations initially drop, and the flow moves towards the laminar state. 
For P1 and P2, this is followed by another reverse transient ($20 < tU_b/D < 50$), during which friction bounces back, and turbulent fluctuations reappear uniformly along the streamwise direction.
Despite identical long-term drag, the two cases diverge immediately after this transient.
For P1, $q$ becomes suddenly concentrated in three coherent spots of activity, which travel at an approximately constant velocity $1.12 U_b$. 
These patches resemble canonical subcritical puffs \citep{avila-barkley-hof-2023}; their properties will be analyzed later in \S\ref{sec:puffAnalysis}.
For P2, turbulence returns to its space-filling state after the transient, with no localization, though intermittency is more pronounced than in the reference case.
The space-time evolution of $q$ for P3 is similar to P1, but with a single turbulent peak instead of three. After a relatively long transient, a single turbulent spot is established, moving at approximately $1.4 U_b$.

The turbulent patches visible in the space--time maps of figure \ref{fig:qxt} resemble the well-documented puffs observed in transitional (canonical) pipe flows. However, $Re_b=4900$ is more than twice the critical value $Re_c=2040$ reported by \cite{avila-etal-2011} for the uncontrolled flow as the critical point for the onset of sustained turbulence. 

Observing localized puffs at such high $Re_b$, along with different flow states (laminar, localized, space-filling) at the same $Re_b$, shows that $Re_b$ alone cannot characterize transition in controlled pipe flow. The same is true for a friction-based Reynolds number: P1 and P2 have the same value of the K\'arm\'an number $R^+$, but exhibit qualitatively different turbulence states (localized turbulence for P1, and space-filling turbulence for P2). 
As discussed for P1 and P3, the flow responds to the control by quickly decreasing friction and turbulent intensity. Before complete laminarization, turbulence bounces back and re-intensifies sufficiently to trigger localized turbulence. The control thus drives the flow into a state that resembles canonical transitional conditions. In this state, space-filling turbulence can no longer be sustained, but survives only in spatially localized patches, where it is so intense that the overall friction is comparable to that of controlled pipes with fully developed (but attenuated) turbulence.
A similar effect has been obtained by \cite{barkley-etal-2015} at a comparable Reynolds number by adding a body force to blunt the shear profile. The initial decrease of $Re_\tau$ observed with StTW suggests that this control also produces a transient phase in which the near-wall shear is dramatically decreased. However, the interaction between the flow and the additional spanwise control gives rise to a velocity profile that, even at fixed $k_x$ and $\omega$, changes with space and time, thus triggering different responses to the flow disturbances, eventually resulting in a different flow state for different forcing parameters. 

\begin{figure}
\centering
\includegraphics[width=0.75\textwidth]{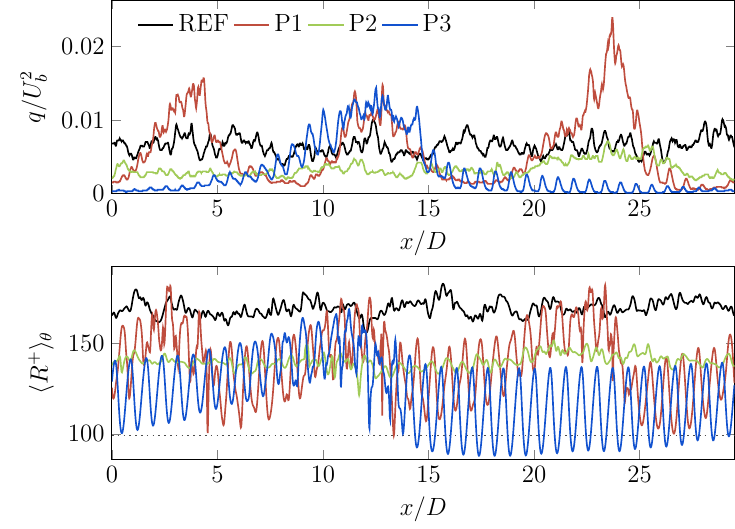}
\caption{Streamwise distribution of $q(x)$ (top) and of the azimuthally averaged wall friction (bottom) for an instantaneous flow field far from the initial transient, for cases REF, P1, P2, and P3. The dotted line on the lower panel is $R^+$ for the laminar flow at $Re_b=4900$.}
\label{fig:qRe}
\end{figure}

The streamwise distribution of $q$ is shown in figure \ref{fig:qRe} for REF, P1, P2, and P3, for a generic flow field extracted at the latest stages of the simulations, when the flow has already assumed its long-term configuration. 
As expected for fully developed turbulence, the reference case shows a nonuniform value of $q$, which randomly oscillates around its mean value. 
The controlled cases differ markedly: P1 and P3 show localized peaks and troughs, while P2 exhibits no localized clustering.
The turbulent clusters in P1 span approximately $5D$.
The single turbulent cluster visible in P3 has an intensity comparable to P1, but a longer streamwise extension ($\sim 10D$), which aligns with information available on structures detected in the canonical case \citep{willis-kerswell-2007, avila-hof-2013}.
The lower panel of  figure \ref{fig:qRe} plots the streamwise evolution of the longitudinal wall friction, quantified via the local value  $\thaver{R^+}(x)$  of the K\'arm\'an number defined in \S \ref{sec:results-dr}. 
Unlike earlier analysis, the control harmonic is retained here to preserve friction modulation. Notably, the modulation is visible only for cases with localized turbulence, whereas P2, with space-filling turbulence, exhibits only small fluctuations around its mean value of $\thaver{R^+} \approx 140$.
In contrast, P1 and P3 with localized turbulence present an evolution of $\thaver{R^+}$ which follows that of $q$ (with a small spatial lag): maxima of $\thaver{R^+}$ correspond to clusters of turbulent activity, and minima fall below the laminar value.

The following sections examine puff properties, using the operator $\paver{\cdot}$, which operates spatial averaging only across the region occupied by a puff. 
First, we must distinguish turbulent from laminar regions.
The axial extent of each puff is extracted from $q(x,t)$ via a thresholding procedure: a position $x$ is considered within a puff when $q(x) > q_t = 3.5 \cdot 10^{-3}U_b^2$. 
Though arbitrary, this threshold satisfies $q_t>q$ in laminar regions, while remaining sufficiently below the maximum in the turbulent core to prevent false positives induced by natural fluctuations; see figure \ref{fig:qRe}. Appendix \ref{sec:qdep} describes how choosing $q_t$ within a reasonable range does not alter the number of puffs and their propagation velocity.
The thresholding criterion is complemented by the requirement that a region with $q > q_t$ must be connected and longer than $L_t = 3D$ to be considered a puff. The value $L_t$ is selected after considering the probability density function (PDF) of the length of the extracted regions with $q>q_t$: as discussed in detail in \S\ref{sec:puffAnalysis}, the PDF is bi-modal, with two evident and separated peaks centered around $L_t$, where the PDF is nearly zero.

\subsection{Characterization}
\label{sec:puffAnalysis}

The turbulent puffs observed in the controlled pipe flow present the same key features of the canonical structures: they maintain their spatial coherence for a long time, are surrounded by quiescent fluid, and travel at a nearly constant speed. 
However, at the present large Reynolds number, no decay of the structures is observed. Indeed, it is known that the decay time increases exponentially with the Reynolds number \citep{hof-etal-2006}, and here $Re_b=4900$ is more than twice the typical values of subcritical uncontrolled pipe flows. At any rate, after the initial transient, puffs do not grow either; besides travelling, they do occasionally interact through merging and splitting, similarly to the uncontrolled puffs observed by \cite{moxey-barkley-2010} at $Re_b \sim 2300$.

\begin{table}
\centering
\begin{tabular}{c|cc|ccccccc}
CASE & $k_x^+$  & $\omega^+$  & $\mathcal{R}$ & $n$ & $L/D$  & $\paver{q}/U_b^2$  & $\paver{R^+}$ & $U_u/U_b$  & $U_d/U_b$  \\ [3pt]
\hline
A    & $6.3 \cdot 10^{-3}$  & $4 \cdot 10^{-2}$ & $40$  & $1$ & $9.9$  & $1.5 \cdot 10^{-2}$ & $150$ & $1.37$  & $1.38$  \\ 
B    & $7.3 \cdot 10^{-3}$  & $4 \cdot 10^{-2}$ & $43$  & $1$ & $9.5$  & $1.3 \cdot 10^{-2}$ & $148$ & $1.40$  & $1.40$  \\ 
C    & $8.4 \cdot 10^{-3}$  & $6 \cdot 10^{-2}$ & $28$  & $3$ & $5.4$  & $1.1 \cdot 10^{-2}$ & $151$ & $1.15$  & $1.15$  \\ 
D    & $9.4 \cdot 10^{-3}$  & $6 \cdot 10^{-2}$ & $35$  & $2$ & $6.2$  & $1.2 \cdot 10^{-2}$ & $150$ & $1.29$  & $1.24$  \\ 
E    & $1.05 \cdot 10^{-2}$ & $6 \cdot 10^{-2}$ & $37$  & $2$ & $6.1$  & $1.1 \cdot 10^{-2}$ & $146$ & $1.25$  & $1.25$  \\ 
F    & $1.15 \cdot 10^{-2}$ & $6 \cdot 10^{-2}$ & $44$  & $1$ & $9.1$  & $1.1 \cdot 10^{-2}$ & $143$ & $1.38$  & $1.35$  \\ 
G    & $1.15 \cdot 10^{-2}$ & $8 \cdot 10^{-2}$ & $27$  & $3$ & $5.7$  & $1.0 \cdot 10^{-3}$ & $144$ & $1.10$  & $1.15$  \\ 
H    & $1.26 \cdot 10^{-2}$ & $6 \cdot 10^{-2}$ & $43$  & $1$ & $8.4$  & $8.7 \cdot 10^{-3}$ & $141$ & $1.37$  & $1.43$  \\ 
I    & $1.26 \cdot 10^{-2}$ & $8 \cdot 10^{-2}$ & $31$  & $3$ & $4.9$  & $9.9 \cdot 10^{-3}$ & $148$ & $1.12$  & $1.13$  \\
L    & $1.29 \cdot 10^{-2}$ & $8.6 \cdot 10^{-2}$ & $29$  & $3$ & $5.5$  & $1.0 \cdot 10^{-2}$ & $149$ & $1.13$  & $1.12$  \\   
M    & $1.36 \cdot 10^{-2}$ & $8 \cdot 10^{-2}$ & $33$  & $3$ & $4.8$  & $8.8 \cdot 10^{-3}$ & $144$ & $1.10$  & $1.11$  \\ 
N    & $1.47 \cdot 10^{-2}$ & $8 \cdot 10^{-2}$ & $36$  & $2$ & $6.3$  & $8.0 \cdot 10^{-3}$ & $143$ & $1.29$  & $1.15$  \\
O    & $-$                  & $-$               & $-$   & $1$ & $5.0$  & $1.1 \cdot 10^{-2}$ & $88$  & $0.96$  & $0.95$  \\                
\end{tabular}
\caption{Characteristics of the turbulent puffs for cases with localized turbulence: the set control parameters $k_x^+$ and $\omega^+$, and the resulting drag reduction rate $\mathcal{R}$, number of puffs $n$ in the domain, length of the puffs $L/D$, cross-plane turbulent kinetic energy $\paver{q}$ and K\'arm\'an number $\paver{R^+}$ averaged within the puff volume, velocities of the upstream and downstream fronts $U_u$ and $U_d$.}
\label{tab:puffChar}
\end{table}

We now provide a quantitative description of the puffs in the controlled pipe flow to ascertain to what extent their features align with those of the canonical puffs and whether they scale in inner or outer units.
Table \ref{tab:puffChar} reports several characteristics, computed over a time window of $t U_b/D=250$, for a subset of cases in the entire database where coherent puffs remain clearly visible throughout the observation period and for which flow and control properties vary smoothly.
This subset is delimited in figure \ref{fig:locTurbMap} with the red contour and different cases represented with dots, whose color encodes the number $n$ of structures present in the domain. 
The table describes each case in terms of its control parameters, the value of $\mathcal{R}$, the number $n$, the average spatial extent $L$ of each puff, and the value of several puff-specific quantities.
Case O (last row) represents canonical transitional puffs. This case is computed by following the procedure described by \cite{moxey-barkley-2010}: starting from a fully developed turbulent field at $Re_b=4900$, the Reynolds number is gradually reduced to $Re_b=2200$, where a single turbulent puff is obtained. Specifically, $Re_b$ is reduced in three discrete steps, with simulations advanced for $tU_b/D=125$ for the first two steps, and $tU_b/D=375$ for the last one (statistics are extracted for the last $tU_b/D=250$ units).

\begin{figure}
\centering
\includegraphics[width=0.75\textwidth]{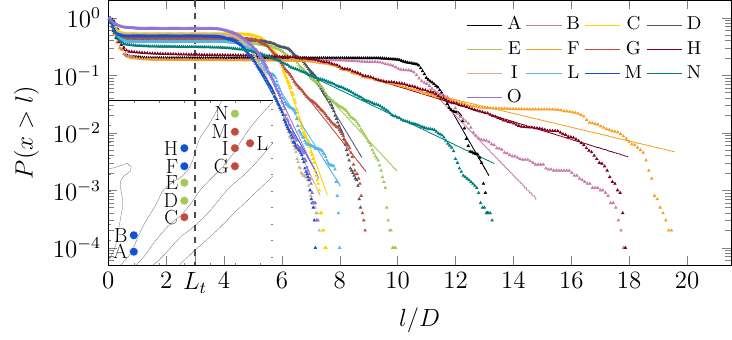}
\includegraphics[width=\textwidth]{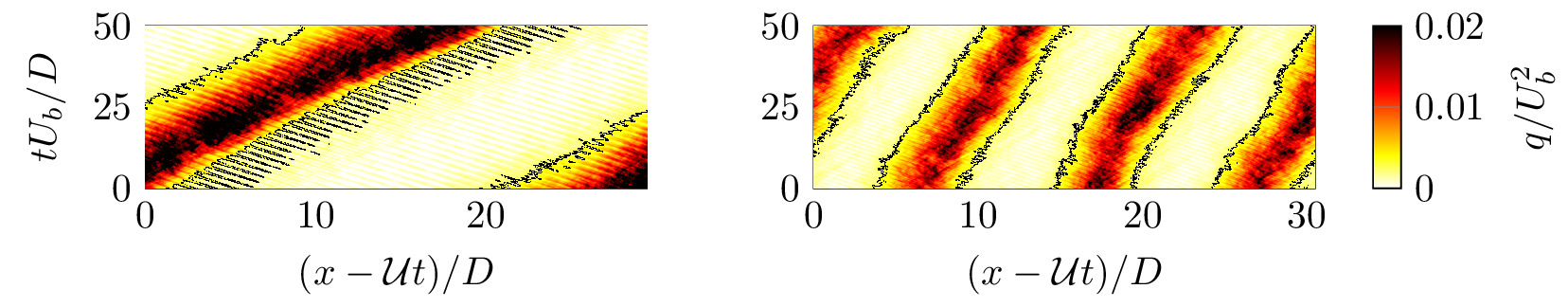}
\caption{Top: cumulative distribution function (CDF) of the puff length for cases listed in table \ref{tab:puffChar}. Continuous lines are the exponential distribution given by equation \ref{eq:CDF}. The inset shows the positions of the different cases within the control parameter space, using the same symbols as in figure \ref{fig:locTurb}. Bottom: space--time evolution of $q(x,t)$ for cases B (left) and I (right). The contour $q=q_t$ marks the spatial extension of the events. $\mathcal{U}=U_b$}
\label{fig:CDF}
\end{figure}

The quantity $L$ in table \ref{tab:puffChar} is the mean streamwise extension of the turbulent region. It is computed by integrating the probability density function of the length $l$ of the extracted turbulent regions for $l>L_t$, $PDF(l>L_t)$, \text{i.e.} only considering the events longer than the threshold length defined in \S\ref{sec:identification}). 
\cite{avila-hof-2013} demonstrated that in pipe flow with localized turbulence, the length of the spatio-temporally intermittent turbulent region exhibits exponential decay beyond a certain threshold $L_0 > L$ (different from the threshold value $L_t$ used to discard small structures). 
This exponential decay characterizes spatio-temporal intermittency, previously reported for other types of flow \citep{chate-manneville-1988, ciliberto-bigazzi-1988}. 
We conduct a similar analysis to distinguish controlled and canonical puffs.
The cumulative distribution function (CDF) of the puff length is first computed. The top panel of figure \ref{fig:CDF} shows the CDF for the cases listed in table \ref{tab:puffChar}. For very small lengths, say $l/D<1$, the CDF suddenly drops from the unitary value, reflecting the presence of short spurious turbulent regions. Then the CDF levels off, indicating that no puffs exist in this range. The threshold value $L_t=3D$ has been selected to always be within the plateau. 
At longer lengths, the CDF decreases again in correspondence with the shortest puffs. The shape of the CDF for all the cases reveals, therefore, very small turbulent regions ($l/D<1$) as well as the longer structures identified as puffs. Turbulent regions of intermediate length are absent, showing a clear bimodal probability density function. 
Moreover, the second decay of the CDF follows an exponential law (a straight line in the plot).
To corroborate this observation, in figure \ref{fig:CDF} the various datasets are accompanied by an exponential distribution
\begin{equation}
CDF(l) = 1 - \mbox{e}^{-(l-L_0)\lambda},
\label{eq:CDF}
\end{equation}
where $L_0$ is the length-scale \citep[larger than the mean length of the puffs, according to][]{avila-hof-2013}) at which the CDF starts to exhibit exponential behaviour, and the decay-rate parameter $\lambda$ is obtained by fitting the data.
The distributions exhibit exponential decay over a wide range of lengths, with gradual deviations appearing only for the longest events. Such deviations were not observed by \cite{avila-hof-2013} and could be related to the confinement of the finite domain, which could impact their statistical distribution, even though the robustness analysis reported in Appendix \ref{sec:domainL} confirms a minor impact of the domain size on the mean properties of the structures.   
Cases B and F stand out and show a distinct behavior, with a substantially slower decay than the exponential fit at intermediate lengths, before the final dropoff at large $l$. To understand this feature, the bottom panel of figure \ref{fig:CDF} compares the space--time evolution of $q(x,t)$ for case B with one case showing a regular exponential decay (case I), showing the contour $q=q_t$. Case B has spikes on the downstream (right) side of the puff, indicating that control is not completely removed by the zeroing of the forced mode. 

While the length is perhaps the most visible quantitative feature of the turbulent puffs, table \ref{tab:puffChar} also contains the values $\paver{q}$ and $\paver{R^+}$ for the cross-plane turbulent kinetic energy $q$ and the local friction Reynolds number $R^+$, averaged within the volume (or the pipe surface, respectively) occupied by the puff. The last columns report the velocities of the upstream and downstream fronts of the puffs, denoted as $U_u$ and $U_d$. 
Most of the characteristics of the controlled puffs resemble those of the uncontrolled ones, including those of case O recomputed here. An exception is the much lower value of $\paver{R^+}$ for case O, which directly ensues from the lower $Re_b$ of the uncontrolled pipe flow. Even though all controlled cases show large drag reduction and therefore a reduced $R^+$, the value of $\paver{R^+}$ remains much higher than that of the canonical puffs, namely $\paver{R^+}=140-150$, compared with the value $\paver{R^+} = 88$ of case O.
Another significant difference lies in the convection speed. The canonical puffs are slower than the bulk flow, with a convection speed of approximately 95\% of the bulk velocity, with $U_u/U_b=0.94$ and $U_d/U_b=0.96$, as reported in the literature and confirmed here for case O. On the contrary, under drag reduction, the puffs are found to travel much faster than $U_b$. For example, case B has $U_u/U_b=1.47$ and $U_d/U_b=1.43$.

\begin{figure}
\includegraphics[width=\textwidth]{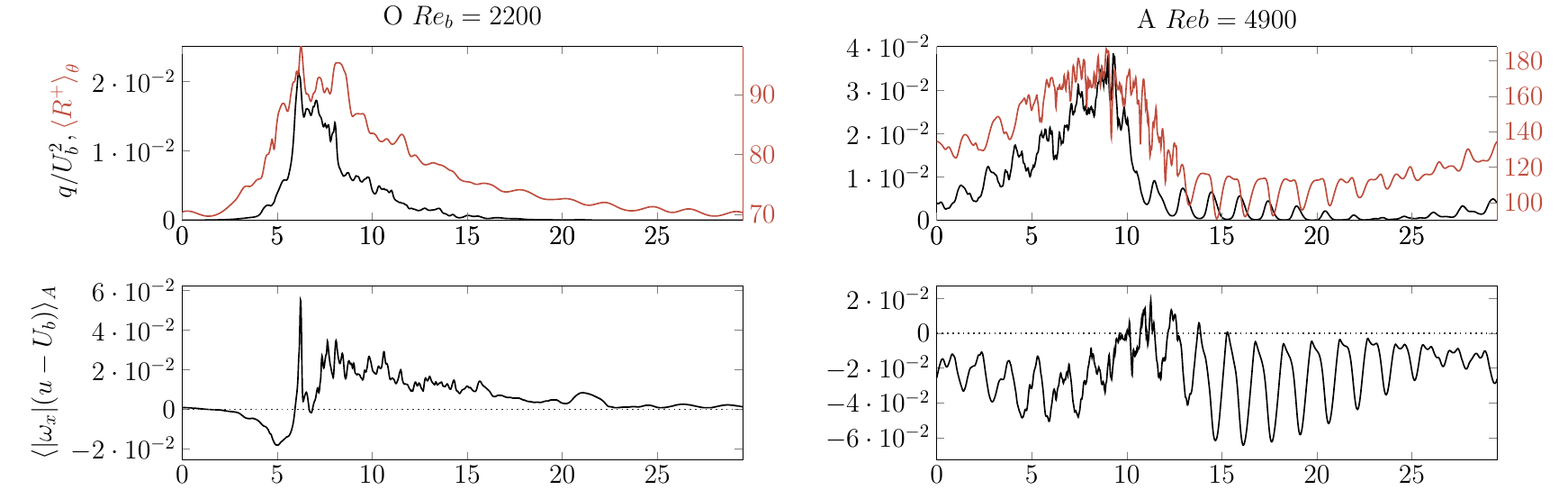}
\caption{Comparison between uncontrolled (case O, left) and controlled (case A, right) puffs. For a single flow field, the top panels plot $\thaver{R^+}$ and $q$ along the pipe at fixed time, and the bottom panels plot the quantity $\aver{|\omega_x|(u-U_b)}_A$.}
\label{fig:puffComp}
\end{figure}

To compare uncontrolled and controlled puffs, figure \ref{fig:puffComp} plots the streamwise evolution of $\thaver{R^+}$ and $q$ for an instantaneous flow field, for cases O and A, both containing a single puff.
Besides the obvious control-induced modulation, some qualitative differences exist in the turbulent regions. In the uncontrolled case, the upstream (left) front presents a steep increase of $q$ from the laminar to the turbulent level, while the downstream (right) front exhibits a more gradual decay to zero. This asymmetry is a known property of the uncontrolled turbulent puffs \citep{hof-etal-2010}. 
The controlled puffs exhibit the opposite behavior, with a gradual rise of $q$ upstream and a steep downstream front.
This dissimilarity is confirmed by the bottom panels of figure \ref{fig:puffComp}, plotting the streamwise evolution of the cross-sectional average of the quantity $\aver{|\omega_x|(u-U_b)}_A$, i.e. the streamwise vorticity (when computing vorticity, the control's harmonic is set to zero for case A) multiplied by the streamwise velocity relative to the bulk velocity.
\cite{hof-etal-2010} introduced this quantity to describe the streamwise rolls generated at the inflection point of the velocity profile in the puff, quantifying the magnitude of the vorticity transport. Here, $\aver{|\omega_x|(u-U_b)}_A$ is negative everywhere for case A, except for a small region at the downstream front, while it is positive everywhere for case O, except for a small region at the upstream front.

\begin{figure}
\includegraphics[width=\textwidth]{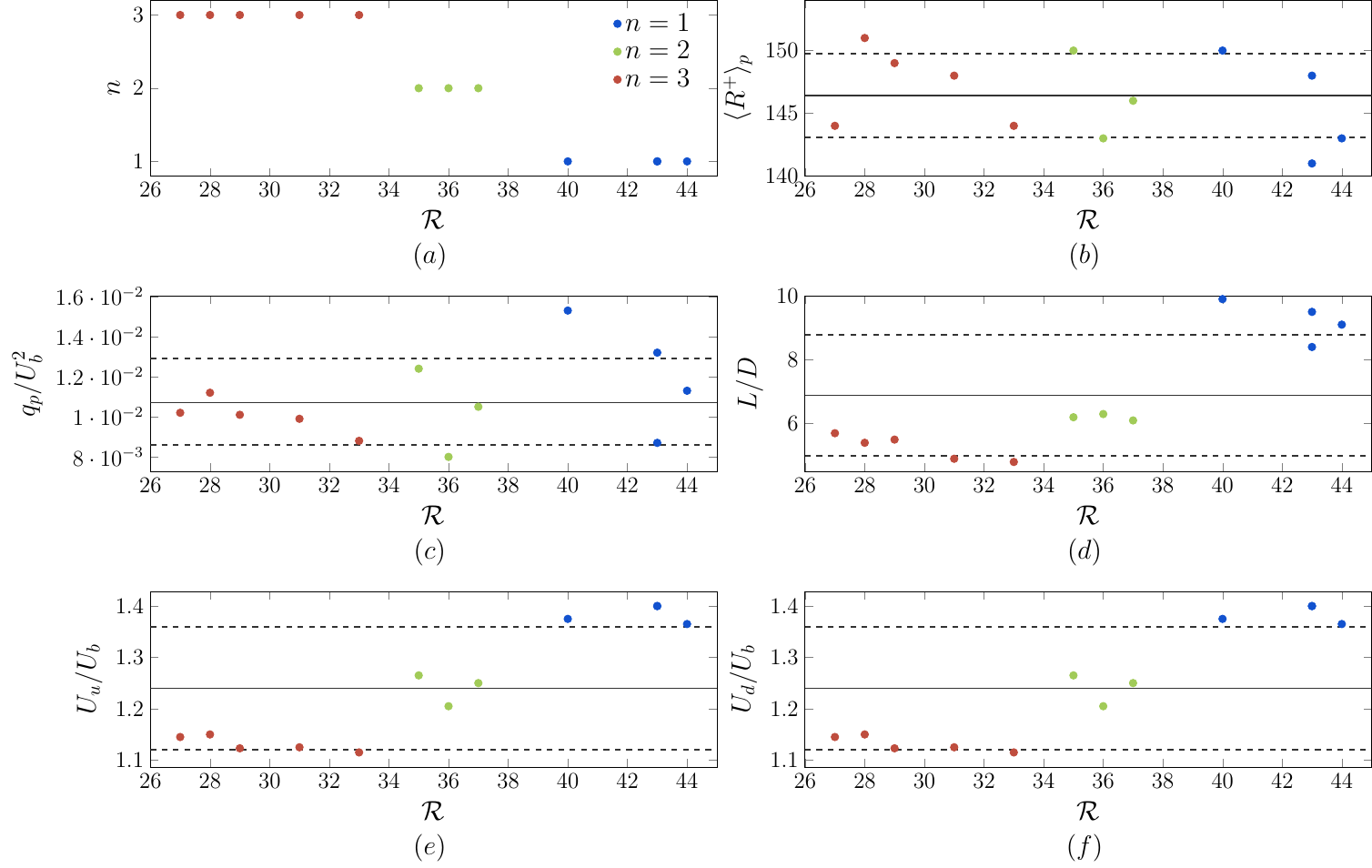}
\caption{Relation between drag reduction rate $\mathcal{R}$ and various quantitative features of the puffs, taken from table \ref{tab:puffChar}. The panels plot: $(a)$, number of puffs in the computational domain; $(b)$, average K\'arm\'an number within the puffs; $(c)$, cross-plane turbulent kinetic energy $\paver{q}$ within the puffs, with inner/outer scaling; $(d)$, puff length; $(e)$-$(f)$,  upstream and downstream front speed. The horizontal solid line indicates the mean value, and the two dashed lines visualize the range of $\pm$ the r.m.s. value.}
\label{fig:RL}
\end{figure}

Figure \ref{fig:RL} illustrates the relation between various quantities extracted from table \ref{tab:puffChar}. 
The number $n$ of puffs observed in the computational domain correlates with $\mathcal{R}$. When $n=1$, drag reduction is large (above $40\%$). For $n=2$,  $35\% < \mathcal{R} < 38\%$, while for $n=3$  $\mathcal{R}$ assumes lower values. 
In the dataset considered here, at most three puffs are observed.
For even smaller $\mathcal{R}$, conventional space-filling turbulence develops and puffs disappear. 

The friction Reynolds number $\paver{R^+}$ within the puff does not correlate with $n$ and $\mathcal{R}$, and hovers in the range 145 -- 150 with a r.m.s. variation of only 2\% of the mean value, indicating that different levels of global drag reduction must derive from the different spatial extent of the laminar region between puffs. The puff intensity, quantified by $\paver{q}/U_b^2$, varies across cases. Expressing it in inner or outer units does not reveal any clear scaling. The r.m.s. variation is about 19\% of the mean value, with a tendency to increase for the lowest $n$. 
In contrast, both the puff length and the upstream and downstream front speeds $U_u$ and $U_d$ decrease with $n$, as seen from panels $(e)$ and $(f)$.
Note that $U_u$ and $U_d$ always take similar values, implying that the puffs neither spread nor merge, but retain their spatial coherency.

Scaling in inner units, i.e. using the friction velocity of the drag-reduced flow $u_\tau$ and viscous length $\delta_\nu$ (superscript $^*$, panels $d$, $f$, and $h$, respectively) does not collapse the data.
For $\paver{q}$ and $U_m$, which increase in outer units with $\mathcal{R}$, inner scaling accentuates these variations. Even for the puff length, which decreases with $\mathcal{R}$, the dispersion of the values around the mean is only slightly reduced. 

Note that typical values of $U_m^*$, where $U_m=(U_u+U_d)/2$, are between $20$ and $30$, i.e., higher than the phase speed of the forcing (maximum value $\sim 6.5$). $U_m^*$ is also larger than the typical convection velocity of near-wall coherent structures in a fully developed wall-bounded turbulent flow ($\sim 10$). 
The puffs therefore substantially differ from canonical near-wall structures, and propagate at velocities that differ significantly from the local bulk speed.
Therefore, although for this subset of cases the level of $\mathcal{R}$ does affect the state of turbulence and links with the number, size, and front convection speed of the puffs, no clear relationship emerges between $\mathcal{R}$ and the turbulence intensity within a puff.

The localized turbulent regions vary widely with $k_x$ and $\omega$. 
While sharing their intermittent character with canonical transitional puffs, they exhibit marked differences, possibly hinting at a distinct physical nature. However, all structures evolve coherently without exhibiting slug-like behavior, splitting, or decay, conversely suggesting a unified underlying mechanism. 
The canonical transitional structures and the localized states investigated here can be considered as members of the same family, whose propertie are dynamically modulated by the control input.

\subsection{Radial structure of the puffs}

Since global features of the puffs exhibit neither inner scaling, where the velocity scale is derived from the wall slope of the axial profile, nor outer scaling, where the velocity scale is determined by the core region, it is of particular interest to examine the radial distribution of the statistical properties of the flow.

\begin{figure}
\includegraphics[width=\textwidth]{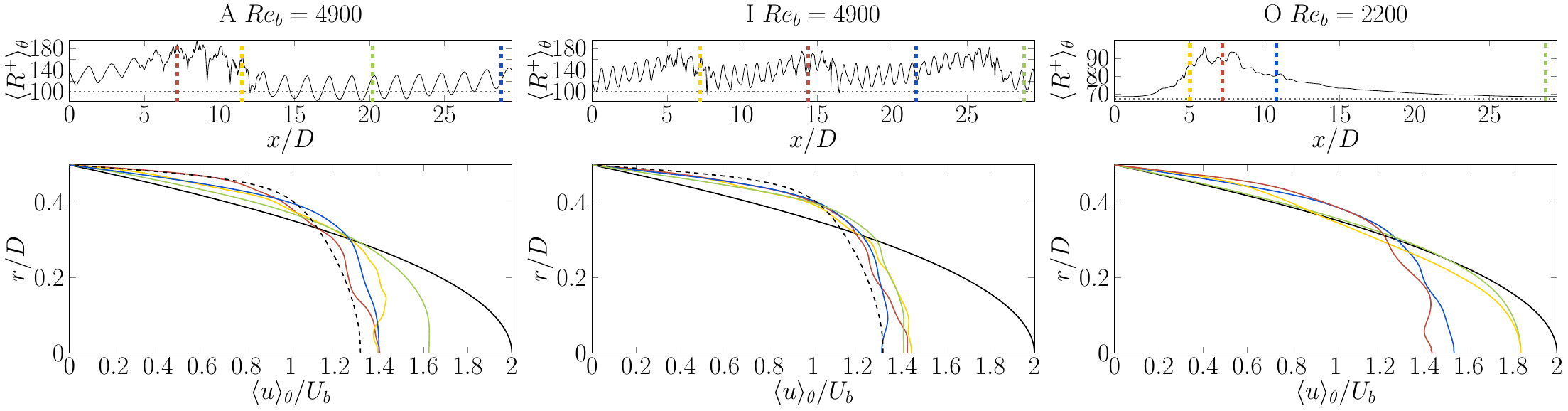}
\caption{Local friction Reynolds number (top) and azimuthally averaged axial velocity profile (bottom) for case A with $n=1$ (left), case I with $n=3$ (center), and the uncontrolled case at $Re_b=2200$ (right). In the top plots, the horizontal dotted line indicates the laminar value. The profiles in the bottom panels are extracted at axial positions marked by vertical dashed lines with matching colors on the top plots. The black solid lines represent the velocity profile for laminar Hagen--Poiseuille flow, while the dashed lines for cases A and I represent the mean velocity profile in the uncontrolled case.}
\label{fig:Umean}
\end{figure}

The instantaneous axial velocity profile is known to exhibit significant variability along the axial direction for canonical puffs \citep{avila-hof-2013}. The axial evolution is shown in figure \ref{fig:Umean}, by considering three cases: case A with $n=1$ and $\mathcal{R}=40\%$ (left), case I with $n=3$ and $\mathcal{R}=31\%$ (center), and the uncontrolled case O with $n=1$ at $Re_b=2200$ (right).
The top panels plot the local $\thaver{R^+}(x)$, after removing the control harmonic.
The bottom panels illustrates the local axial velocity profile $\thaver{u}(r)$, extracted from an instantaneous snapshot, azimuthally averaged, and plotted at different axial locations representing the mild (blue) and steep (yellow) fronts of the puff. Profiles extracted from the region of most intense turbulent activity (red) and within the laminar plateau (green) are also shown. The laminar parabolic profile at the same flow rate (black) is plotted for reference, together with the uncontrolled turbulent velocity profile for cases at $Re_b=4900$.

When $n=1$ (case A, left), the local friction within the turbulent spot reaches a maximum of $\thaver{R^+} \approx 180$, which is slightly higher than that of the uncontrolled pipe flow at the same $Re_b$. The corresponding velocity profile (red), which exhibits slight oscillations because only azimuthal averaging is performed, resembles a typical turbulent profile, especially in the vicinity of the wall, with a nearly flat core and a steep near-wall gradient. 
Downstream of the position of maximum $R^+$, a recovery region is observed where the local friction decreases. The velocity profile (yellow) approaches the parabolic one: the flat central region is still discernible but becomes narrower, and near the wall, the profile is quite close to the parabolic one.
The laminar value of $R^+$ is reached at about $x/D \approx 20$, but the corresponding profile  (green) does not fully recover the laminar profile.
It still exhibits a rather flat bulk region with a centerline velocity of $\approx 1.6 U_b$ (instead of the laminar value $2 U_b$). For $r/D$ between 0.3 and 0.5, the actual velocity exceeds the laminar value.
Lastly, upstream of the plateau, the front region shows a progressive increase in $R^+$ up to its peak. The velocity profile (blue) approaches the turbulent one again, although with lower velocities near the wall.

When $n=3$ (case I, center), the separation between adjacent turbulent regions is limited: the downstream side of one puff penetrates the upstream side of the next one. As a result, the local friction remains significantly higher than the laminar value across the entire length of the pipe. In contrast to the $n=1$ case, the velocity profile exhibits minimal variations throughout the domain, including at the centerline. Even at the location where the local wall shear stress reaches its minimum, the velocity profile (green) remains similar to the others and quite far from the parabola.

The single puff at $Re_b=2200$ (case O, right) is similar to the case with $n=1$. The size of the turbulent puff is comparable, and the expected sharp upstream and mild downstream fronts are observed. Interestingly, the upstream (yellow) profile presents an inflection point similar to the one observed by \cite{hof-etal-2010}. A closer look at the corresponding profiles for $n=1$ (at $x/D=11.5$) and $n=3$ (at $x/D=7.2$) suggests the presence of similar, although less evident, inflection points located at comparable wall distances.

\begin{figure}
\includegraphics[width=\textwidth]{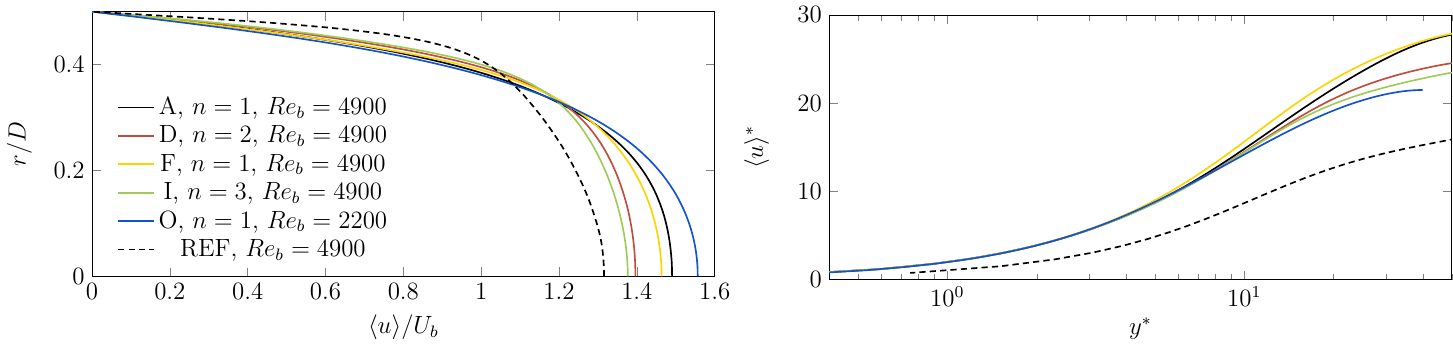}
\includegraphics[width=\textwidth]{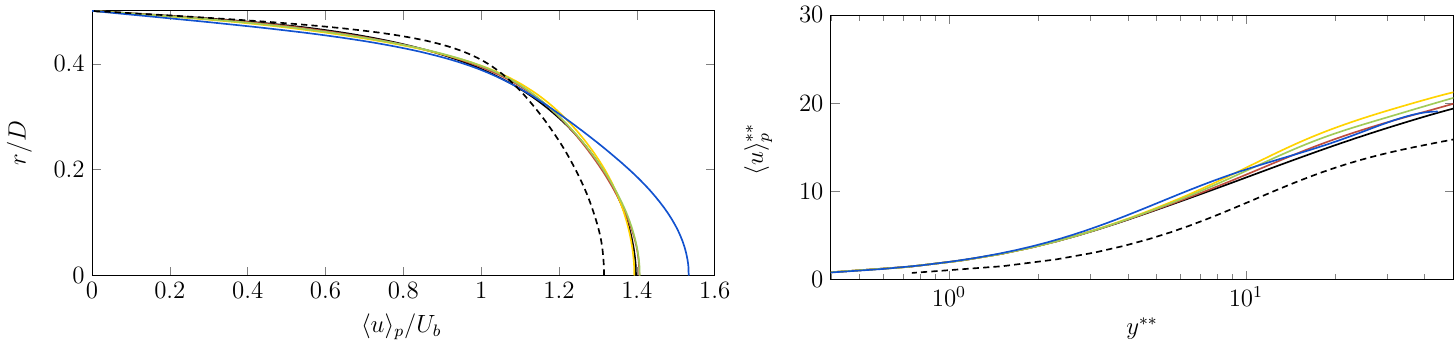}
\caption{Mean velocity profiles in outer (left) and actual inner (right) scaling. Top: global mean profile $\aver{u}(r)$; bottom: mean profile $\paver{u}(r)$ averaged only over the puffs. The dashed lines represent the mean velocity profile for the uncontrolled case at $Re_b=4900$.}
\label{fig:UmeanComparison}
\end{figure}

Figure \ref{fig:UmeanComparison} plots mean velocity profiles, normalized in inner and outer units, and compares different criteria for averaging. The top panels present the global mean $\aver{u}(r)$, and the bottom ones show the puff-averaged profile $\paver{u}(r)$.
Four controlled cases (namely A, D, F, and I) with different values of $n$ are considered and compared with the uncontrolled case O at $Re_b=2200$: two cases have $n=1$, while exhibiting qualitatively different puffs in terms of intensity and length; one case has $n=2$ and one has $n=3$. The mean turbulent velocity profile for the uncontrolled case at $Re_b=4900$ is added as a reference.

The profile $\aver{u}(r)$ in outer scaling (top left) varies across the different cases, being the centerline velocity $\aver{u}(0)$ largest for the uncontrolled case O, and decreasing with increasing $n$ (and therefore with decreasing drag reduction) and reaching the minimum for the fully developed turbulent profile. Note, however, that the centerline velocity is not determined solely by $\mathcal{R}$, as $\aver{u}(0)$ is higher for case A with $\mathcal{R}=40$ than for case F with $\mathcal{R}=44$. 
The right plot shows the same profiles in the law-of-the-wall form with actual viscous scaling defined in \S \ref{sec:sims}. The controlled cases collapse, as expected, near the wall, whereas they are quite scattered in the incipient logarithmic region, barely visible because of the low $Re$, with case F above O and the others below. All the controlled profiles show an upward shift of the logarithmic region compared to REF, in agreement with the drag-reducing action of the control.

Despite the remarkable differences in the spatial arrangement of the puffs, all the puff-averaged profiles $\paver{u}(r)$ collapse in external units.
The profile for the uncontrolled case O stands out. A lower wall gradient and, therefore, lower velocity levels in the near-wall region are expected, since $\paver{R^+}$ is smaller (see table \ref{tab:puffChar}). Therefore, since the flow rate is the same by design, velocities must be larger in the core. 
The maximum centerline velocity for O reaches $\paver{u}(0)/U_b \sim 1.53$, which is below the laminar value $\aver{u}(0)/U_b =2$. For the controlled cases, a value $\paver{u}(0)/U_b \sim 1.4$ is observed.
When the friction velocity is computed locally within the puff, such local inner scaling leads to the expected collapse of the various profiles in the near-wall region. However, owing to the limited variability of $\paver{R^+}$, the velocity profiles (including the uncontrolled case O) collapse quite well in the logarithmic region too.

\begin{figure}
\includegraphics[width=\textwidth]{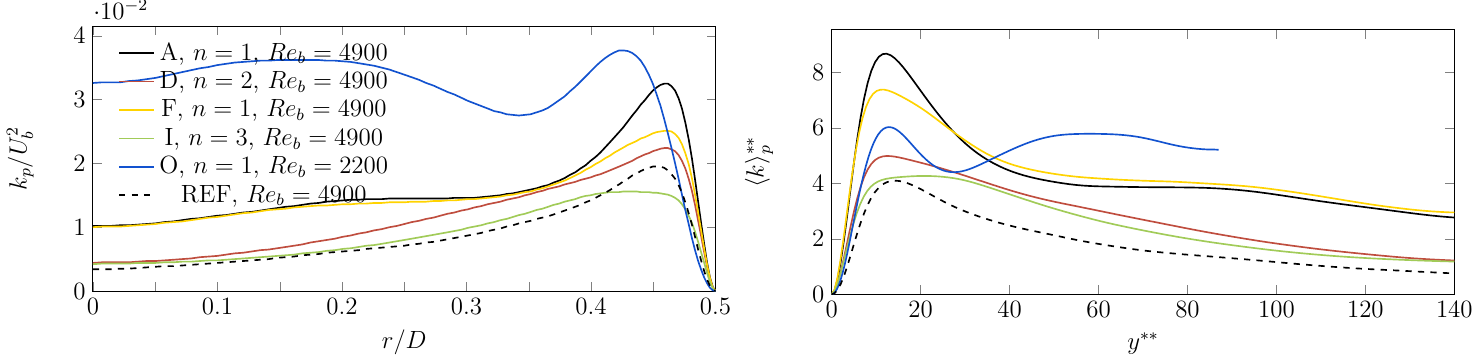}
\caption{Radial distribution of the puff-averaged turbulent kinetic energy $\paver{k}$, scaled in external (left) and actual  (right) viscous units. The dashed lines represent the turbulent kinetic energy profile for the uncontrolled case at $Re_b=4900$.}
\label{fig:kRadial}
\end{figure}
 
Figure \ref{fig:kRadial} plots the radial distribution of the puff-averaged turbulent kinetic energy $\paver{k}$ (after removal of the forced mode), scaled in external units (left) and actual local viscous units (right). This quantity differs from $\paver{q}$ since it includes fluctuations in the streamwise direction, in addition to a 0.5  prefactor. Also in this figure, the turbulent kinetic energy profile for the uncontrolled case at $Re_b=4900$ is added as a reference.
In general, apart from case F, when scaled in external units, the puffs are more energetic than the fully developed turbulent uncontrolled flow.
The shape of the $\paver{k}$ profile for the uncontrolled case O presents two peaks of comparable intensity: one is close to the wall and a second, broader, towards the centerline.
The controlled profiles are qualitatively different from O. The position of the near-wall peak of $\paver{k}$ is variable in outer units, but collapses to the canonical value of about 15 wall units when scaled in local viscous units.
Looking at the shape of the profiles, two groups of curves can be identified. For cases A, F (and B, which is not shown), the turbulent kinetic energy peaks near the wall, then drops and flattens out  up to far from the wall ($r/D \sim 0.2$, $y^* \sim 100$). This plateau is absent for the other cases, independently of the value of $n$.
Note that the structures associated with cases A and F are more energetic than the others, despite the level of drag reduction being substantially higher. This can be explained by the different number of structures $n$: for cases A and F, $n=1$, meaning that despite being the structures more energetic, the overall friction is reduced due to the extended laminar or almost laminar recovery regions observed, for example, in figure \ref{fig:qRe}.

The previous observations provide insight into the internal structure of the controlled puffs. Their faster convection can in principle be related to faster near-wall structures, which shift in the wall-normal direction under StTW \citep{gallorini-quadrio-gatti-2022}. However, the unchanged position of the peak in $\paver{k}$, shown in figure \ref{fig:kRadial}, rules out such a possibility.
By contrast, the canonical case O at low $Re$ presents a peak of $\paver{k}$ near the center. In this case, the convective velocity of the puffs actually decreases (see table \ref{tab:puffChar}), indicating that the radial distribution of turbulent kinetic energy does not determine the propagation velocity of the puffs.
 \section{Concluding discussion}
\label{sec:conclusion}

This work has studied the effect of streamwise-travelling waves of azimuthal velocity on a turbulent pipe flow at $Re_b=4900$ using direct numerical simulations. We have provided the first comprehensive drag reduction map for the pipe geometry, and investigated the subcritical turbulent state that develops in the controlled pipe.

Considering a pipe of length $30D$, we have examined how control parameters $k_x$ and $\omega$ affect drag reduction at fixed amplitude $A^+=14.5$, including large-wavelength configurations. A drag reduction map analogous to that available for plane channel flow \citep{quadrio-ricco-viotti-2009} has been obtained. 
The comparison between the two geometries has highlighted that the pipe outperforms the channel in terms of maximum drag reduction $\mathcal{R}$ and net power saving $\mathcal{S}$. Notably, the controlled pipe relaminarizes, in contrast to the channel flow. However, the channel performs slightly better for several non-optimal combinations of $k_x$ and $\omega$.

Focusing on the turbulence state created by the control parameters, we have observed localized turbulent spots embedded in quiescent fluid; they resemble the puffs typical of the subcritical transition in an uncontrolled pipe flow. However, a crucial difference exists: in canonical pipe flow transition, turbulence becomes sustained and space-filling above a Reynolds number of $Re_b=2600$ \citep{moxey-barkley-2010}, whereas the present Reynolds number is much higher at $Re_b=4900$.
Since wall-based control alters the relation between $Re_b$ and $Re_\tau$ (or equivalently, the Kármán number $R^+$ for pipe flow), one might expect $Re_\tau$ to better describe the transition process. However, cases with identical $R^+$ but different controls exhibit different turbulent states. This demonstrates that $Re_\tau$ alone cannot characterize transition in the controlled flow; the control parameters play an essential role alongside $Re_b$.

A criterion based on local variance of velocity fluctuations has been introduced to discriminate spatially localized turbulence, enabling the identification of a subset of cases used to quantitatively study turbulent puffs.
We have compared these puffs with those observed in the canonical pipe. Their lengths are comparable, but other properties differ significantly. Interestingly, uncontrolled puffs travel slower than $U_b$, whereas the controlled structures exceed $U_b$, with convection speed showing weak dependence on drag reduction.

In conclusion, StTW modifies pipe flow beyond the pure effect on the friction drag: it fundamentally alters transition onset and the structure of turbulence.
Under high drag reduction, turbulence organizes itself into localized patterns reminiscent of the uncontrolled flow at much lower $Re$. Neither $Re_b$ nor $Re_\tau$ alone fully characterizes the flow regime or predicts transition. 

Future works should assess the generality of these findings across parameter space, examining the physical mechanisms underlying laminarization and localized turbulence.
Whether this behavior is characteristic of pipe flow alone also remains unclear. To our knowledge, similar localized turbulence has not been observed for plane channel flow. However, in subcritical conditions, intermittent laminar-turbulent patterns appear in plane Poiseuille flow and other canonical flows, including plane Couette and Taylor--Couette configurations \citep{tuckerman-chantry-barkley-2020}.
Interesting questions are whether such intermittent patterns appear when control is applied to these other geometries, and whether puff-like structures comparable to those observed here exist in geometries with similar transition behavior \citep[e.g., the square duct][]{takeishi-etal-2015}.

These results emphasize the importance of understanding how both active approaches, such as StTW, and passive techniques, like riblets or surface roughness, modify the turbulent state in the subcritical regime.
 
\appendix
\section{Domain length and initial conditions}
\label{sec:domain}

Experiments on turbulent puffs typically require extremely long domains to fully characterize the process \citep{avila-willis-hof-2010}.
To keep the computational cost of this parametric study reasonable, the streamwise length of the pipe is fixed for the bulk of the numerical study at $L_x \simeq 30D$. The finite length of the periodic domain imposes confinement. The number of puffs that exist at a given time in a periodic domain is quantized to integer values. Initial conditions may also influence the flow state. Therefore, here we quantify how flow statistics are influenced by the domain length \citep[as done, for example, by][]{chin-etal-2010} and by the initial conditions.

\subsection{Effect of the domain length}
\label{sec:domainL}
To explore quantization effects, additional simulations are carried out. One case with localized turbulence ($k_x^+=0.0129$, $\omega^+=0.086$) is selected, and the pipe length is varied between $8.5D$ and $122D$, encompassing the baseline $L_x=30D$ used in the main study. All other simulation parameters, including the spatial resolution, remain unchanged.

\begin{table}
\centering
\begin{tabular}{c|ccccccc}
$L_x/D$ & $8.5$  & $16$ & $24.5$ & $32$ & $39$ & $47.5$ & $122.5$\\ [3pt]
\hline
$\mathcal{R}$  & $27.9$  & $27.5$ & $27.8$ & $29.1$ & $27.5$ & $27.7$ & $28.7$ \\
$\delta \mathcal{R}$ & $0.53$ & $0.47$  & $0.34$ & $0.44$ & $0.32$ & $ 0.28$ &  $0.31$ \\
$n$            & $1$     & $2$    & $3$    & $3$    & $5$    & $6$    & $13$ \\
$L/D$          & $4.9$   & $4.6$ & $5.11$ & $5.3$ & $5.4$ & $4.6$ & $5.1$ \\
$\paver{q}/U_b^2$          & $9.1\cdot 10^{-3}$  & $9.3\cdot 10^{-3}$ & $8.5\cdot 10^{-3}$ & $1\cdot 10^{-2}$ & $8.2\cdot 10^{-3}$ & $9.1\cdot 10^{-3}$ & $0.98\cdot 10^{-3}$  \\
$U_m/U_b$          & $1.11$  & $1.11$ & $1.11$ & $1.13$ & $1.02$ & $1.10$ & $1.09$ \\
$nD/L_x$        & $0.116$  & $0.126$ & $0.122$ & $0.095$ & $0.128$ & $0.126$ & $0.106$ 
\end{tabular}
\caption{Dependence of $\mathcal{R}$, its uncertainty $\delta \mathcal{R}$, and characteristics of the turbulent regions (number $n$, length $L$, cross-plane turbulent kinetic energy $\paver{q}$, mean puff velocity $U_m$ and number of puffs per unit length $nD/L_x$) with the domain length for the localized-turbulence case with control parameters $k_x^+=0.0129$ and $\omega^+=0.086$.}
\label{tab:domainDR}
\end{table}

Results are summarized in table \ref{tab:domainDR} in terms of $\mathcal{R}$ and its uncertainty $\delta \mathcal{R}$, computed as in \cite{gatti-quadrio-2016}:
\begin{equation}
\delta \mathcal{R} = 100 \cdot\frac{C_f}{C_{f,0}}\sqrt{\Bigl(\frac{\delta C_f}{C_{f}}\Bigr)^2+\Bigl(\frac{\delta C_{f,0}}{C_{f,0}}\Bigr)^2},
\end{equation}
where $\delta C_{f}$ and $\delta C_{f,0}$ are estimated using the procedure proposed by \cite{russo-luchini-2017}. The relatively small changes of $\mathcal{R}$ with the domain length, together with the small uncertainties, suggest that the confinement effect does not significantly alter the extent of turbulent regions and, consequently, the value of drag reduction.

\begin{figure}
\includegraphics[width=\textwidth]{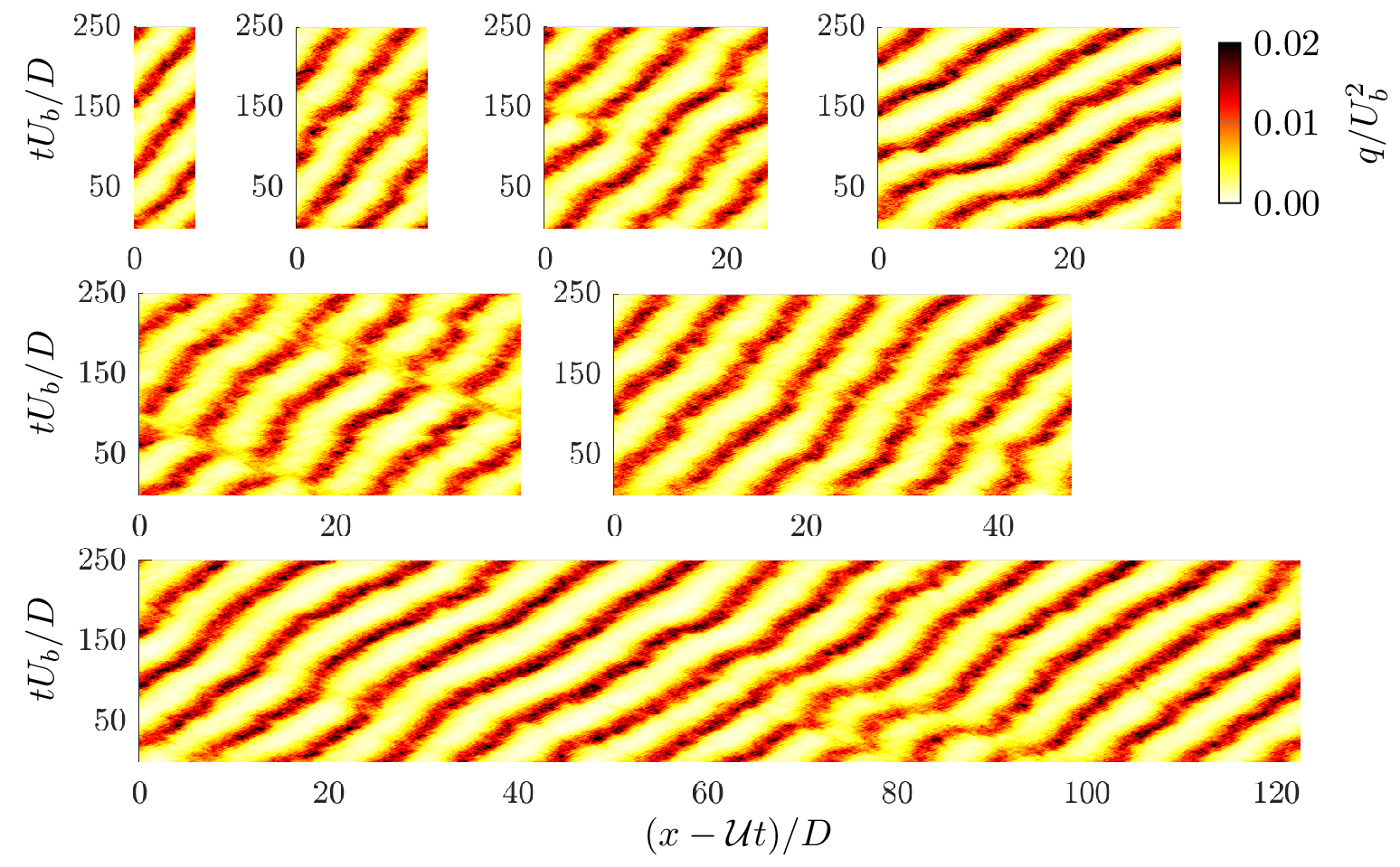}
\caption{Space--time evolution of $q(x,t)$ (for the case with $k_x^+=0.0129$ and $\omega^+=0.086$) for different $L_x$: $8.5D$, $15D$, $24.5D$, $32D$ (top panel), $39D$, $47.5D$ (central panel) and $122.5D$ (bottom panel).}
\label{fig:domainDep}
\end{figure}

Figure \ref{fig:domainDep} compares the space--time evolution of $q(x,t)$, as in figure \ref{fig:qxt} for the baseline cases. For consistency, $\mathcal{U}=U_b$ for all the cases. As expected, the number $n$ of turbulent patches varies with the domain length, but the relationship between $n$ and $L_x$ is not straightforward. 
In the shortest domain with $L_x = 8.5D$, a single puff exists within the pipe. Doubling to $L_x = 16 D$ produces $n=2$. For $L_x=24.5D$, $n=3$ puffs are observed. As the domain length further increases, the number increases non-linearly: $n=3$ for $L_x=32D$, $n=5$ at $L_x=39D$, $n=6$ at $L_x=47.5D$, up to $n=13$ for the longest domain with $L_x=122.5D$.

Figure \ref{fig:domainDep} confirms that several features of the puffs are qualitatively affected by the domain length. In pipes with $L_x = 16D$ and $L_x = 47.5D$, the puffs are similar in length and intensity. In contrast, for $L_x = 32D$ they show greater extension and higher intensity, as quantified in Table \ref{tab:domainDR}, which summarizes the puff characteristics. In particular, the table reports the number $n$ of puffs, their length $L/D$, their cross-plane turbulent kinetic energy $\paver{q}$, their mean convecting speed $U_m$, defined as the average of the upstream and downstream front velocities, and their number per unit length $n D / L_x$.

Overall, the domain length produces only minor variations in puff arrangement and intensity. The consistent drag reduction values across all investigated scales confirm that confinement effects are negligible for the global flow behavior.

\subsection{Effect of the initial conditions}

\begin{figure}
\centering
\includegraphics[width=\textwidth]{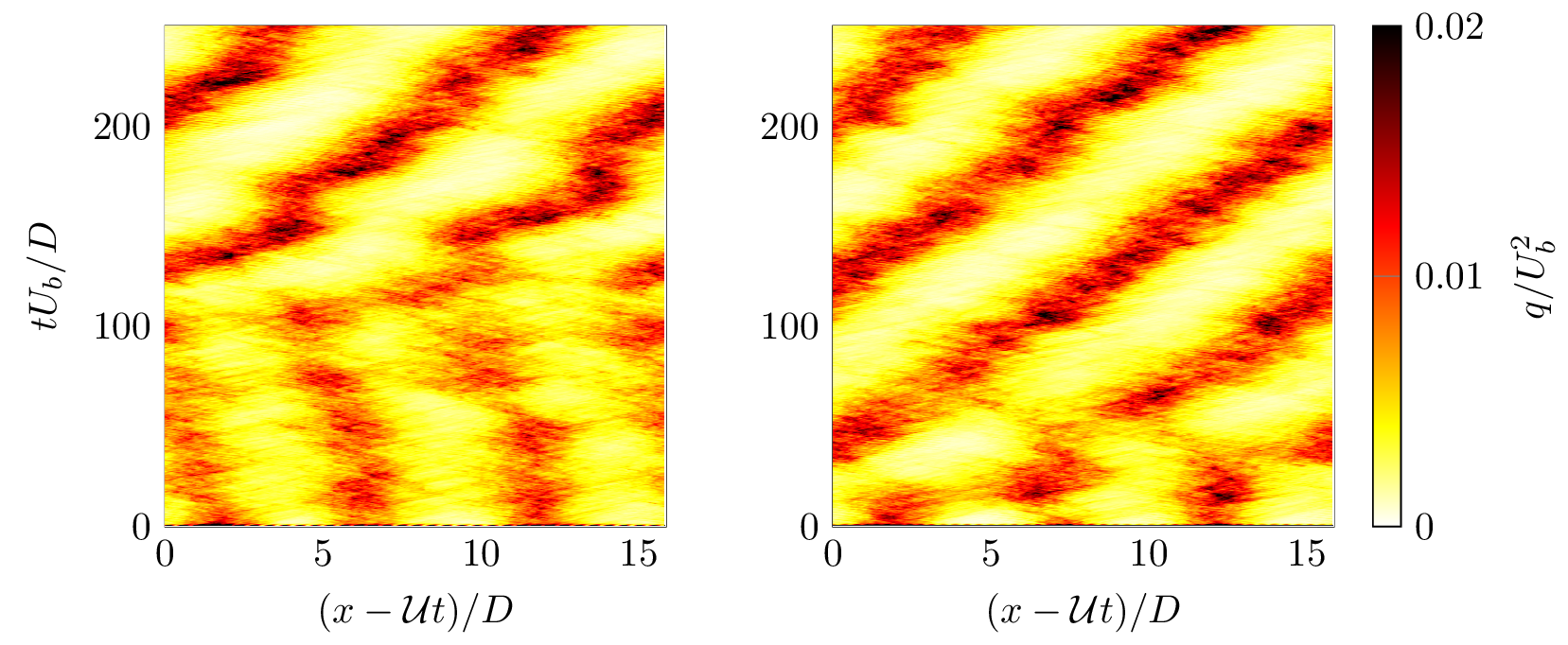}
\caption{Space--time evolution of $q(x,t)$ (for the case with $k_x^+=0.0129$ and $\omega^+=0.086$) and $L_x= 16D$ with initial condition originating from  $L_x= 24.5D$ (left) and $L_x= 32D$ (right). }
\label{fig:domainDep2}
\end{figure}

To test dependence on initial conditions, we performed two additional simulations at $L_x = 16D$, where $n=2$ is expected (figure \ref{fig:domainDep}). Initial velocity fields were taken from simulations at $L_x=24.5D$ and $L_x=32D$ (both with $n=3$), then interpolated onto the shorter domain while preserving the Fourier mode count. This creates a non-equilibrium initial condition that must adjust.
Figure \ref{fig:domainDep2} shows $q(x,t)$ for the first $275U_b/D$ time units, with $\mathcal{U}=U_b$.

For both initial conditions, the number of structures in the domain eventually decreases from $n=3$ to $n=2$, indicating that the final turbulent state does not depend on the initial conditions for this case. 
The main statistical features remain nearly unaffected: the first case yields $\mathcal{R}=27.7 \pm 0.4$ and the second $\mathcal{R}=27.0 \pm 0.4$, compared to the reference value $\mathcal{R}=27.5 \pm 0.5$ (\S\ref{sec:domainL}).
The transient behavior differs, however. The simulation initialized with the shorter-domain solution exhibits a longer adjustment period during which three puffs persist, traveling at speeds below $U_b$. At $t \approx 130$, rapid transition to the $n=2$ state occurs via coalescence, and the resulting structures immediately propagate faster than $U_b$.
In contrast, the simulation from the longer-domain solution shows a much shorter transient with no evidence of structures moving slower than $U_b$.
 \section{Effect of the threshold value $q_t$ on the quantitative features of the puffs}
\label{sec:qdep}

The characteristics of the puffs are affected by the choice of the threshold value $q_t$ for the cross-sectional kinetic energy $q$, used to identify turbulent regions. Table \ref{tab:puffChar2} presents the same puff characteristics as Table \ref{tab:puffChar}, computed for different threshold values. We consider $q_t = 7 \cdot 10^{-3}U_b^2$ (twice the baseline threshold), as well as a case-dependent threshold set at 10\% of the local maximum $q$.
As expected, $L$, $q_p$, and $R^+_p$ show variations with $q_t$, whereas $U_u$ and $U_d$ remain nearly constant. Front velocities exhibit the same correlation with $\mathcal{R}$, and all findings from \S\ref{sec:puffAnalysis} are supported.

\begin{table}
\centering
\begin{tabular}{c|ccccc}
CASE & $L/D$  & $\paver{q}/U_b^2$  & $\paver{R^+}$ & $U_u/U_b$  & $U_d/U_b$  \\ 
\hline
A    & $7.3$  & $1.8 \cdot 10^{-2}$ & $155$ & $1.39$  & $1.40$  \\ 
B    & $7.0$  & $1.6 \cdot 10^{-2}$ & $152$ & $1.35$  & $1.36$  \\ 
C    & $3.7$  & $1.4 \cdot 10^{-2}$ & $157$ & $1.15$  & $1.15$  \\ 
D    & $4.4$  & $1.5 \cdot 10^{-2}$ & $154$ & $1.28$  & $1.23$  \\ 
E    & $4.1$  & $1.3 \cdot 10^{-2}$ & $150$ & $1.17$  & $1.20$  \\ 
F    & $6.1$  & $1.4 \cdot 10^{-2}$ & $147$ & $1.31$  & $1.33$  \\ 
G    & $3.7$  & $1.3 \cdot 10^{-2}$ & $154$ & $1.12$  & $1.11$  \\ 
H    & $4.6$  & $1.2 \cdot 10^{-2}$ & $145$ & $1.35$  & $1.40$  \\ 
I    & $3.2$  & $1.2 \cdot 10^{-2}$ & $152$ & $1.13$  & $1.11$  \\
L    & $3.7$  & $1.3 \cdot 10^{-2}$ & $149$ & $1.13$  & $1.11$  \\               
M    & $2.9$  & $1.1 \cdot 10^{-2}$ & $149$ & $1.11$  & $1.11$  \\ 
N    & $3.9$  & $1.0 \cdot 10^{-2}$ & $147$ & $1.26$  & $1.18$  \\ 

\end{tabular}
\begin{tabular}{c|ccccc}
CASE & $L/D$  & $\paver{q}/U_b^2$  & $\paver{R^+}$ & $U_u/U_b$  & $U_d/U_b$  \\ 
\hline
A    & $8.3$  & $1.7 \cdot 10^{-2}$ & $151$ & $1.39$  & $1.38$  \\ 
B    & $9.1$  & $1.4 \cdot 10^{-2}$ & $148$ & $1.43$  & $1.44$  \\ 
C    & $5.1$  & $1.2 \cdot 10^{-2}$ & $152$ & $1.13$  & $1.12$  \\ 
D    & $6.0$  & $1.3 \cdot 10^{-2}$ & $151$ & $1.25$  & $1.26$  \\ 
E    & $6.3$  & $1.0 \cdot 10^{-2}$ & $146$ & $1.23$  & $1.20$  \\ 
F    & $8.2$  & $1.2 \cdot 10^{-2}$ & $143$ & $1.32$  & $1.38$  \\ 
G    & $5.8$  & $1.0 \cdot 10^{-2}$ & $152$ & $1.16$  & $1.12$  \\ 
H    & $8.6$  & $8.6 \cdot 10^{-3}$ & $138$ & $1.37$  & $1.42$  \\ 
I    & $5.2$  & $9.8 \cdot 10^{-3}$ & $147$ & $1.12$  & $1.11$  \\ 
L    & $5.4$  & $1.0 \cdot 10^{-2}$ & $152$ & $1.13$  & $1.12$  \\                            
M    & $5.3$  & $8.3 \cdot 10^{-3}$ & $142$ & $1.10$  & $1.11$  \\ 
N    & $7.6$  & $7.2 \cdot 10^{-3}$ & $141$ & $1.15$  & $1.22$  \\
\end{tabular}

\caption{Quantitative characteristics of the localized turbulent regions for various  threshold values $q_t$ for the cross-plane turbulent kinetic energy. Left: $q_t=7 \cdot 10^{-3}U_b^2$. Right: $q_t=0.1q_{max}$.}
\label{tab:puffChar2}
\end{table}

An additional consideration concerns the sensitivity of $q_t$ to $Re_b$, discussed in \S\ref{sec:puffAnalysis}. Table \ref{tab:puffChar} shows that case O at $Re_b=2200$ contains shorter structures than case A, despite both having $n=1$.
Table \ref{tab:puffChar3} therefore reports puff properties for case O with a reduced threshold $q_t = 2 \cdot 10^{-3}U_b^2$, preserving the same ratio of $q_t$ to peak kinetic energy as case A. The reduced $q_t$ yields lower $\paver{q}/U_b^2$ and $\paver{R^+}$ compared to Table \ref{tab:puffChar}, while $L/D$ increases. Again, front speeds are nearly insensitive to $q_t$. These results confirm that puffs in case O are shorter, weaker, and slower than the $n=1$ cases at $Re_b=4900$, regardless of threshold choice.

\begin{table}
\centering
\begin{tabular}{c|ccccc}
CASE & $L/D$  & $\paver{q}/U_b^2$  & $\paver{R^+}$ & $U_u/U_b$  & $U_d/U_b$  \\ 
\hline
O    & $6.5$  & $9.2 \cdot 10^{-3}$ & $87$ & $0.96$  & $0.95$  \\ 
\end{tabular}
\caption{Quantitative characteristics of the localized turbulent regions in case O for threshold value for $q_t=2 \cdot 10^{-3}U_b^2$.}
\label{tab:puffChar3}
\end{table}

\section*{Declaration of Interests} The authors report no conflict of interest.

\section*{Acknowledgments} 
Simulations were performed thanks to the Swedish National Infrastructure for Computing (SNIC) at the NSC (Linköping University). Computing time has also been provided by the Italian supercomputing center CINECA under the ISCRA C project DWaves. DM acknowledges the financial support provided by the Knut and Alice Wallenberg Foundation. Preliminary results of this research were presented by EG at the European Drag Reduction and Flow Control Meeting (EDRFCM 2024) in Turin, Italy \citep{gallorini-etal-2024}. 

\bibliographystyle{jfm} 
\bibliography{./Wallturb}

@article{auteri-etal-2010,
  title = {Experimental Assessment of Drag Reduction by Traveling Waves in a Turbulent Pipe Flow},
  author = {Auteri, F. and Baron, A. and Belan, M. and Campanardi, G. and Quadrio, M.},
  year = 2010,
  journal = {Phys. Fluids},
  volume = {22},
  number = {11},
  pages = {115103/14}
}

@article{avila-barkley-hof-2023,
  title = {Transition to {{Turbulence}} in {{Pipe Flow}}},
  author = {Avila, M. and Barkley, D. and Hof, B.},
  year = 2023,
  journal = {Ann. Rev. Fluid Mech.},
  volume = {55},
  number = {1},
  pages = {575--602},
  doi = {10.1146/annurev-fluid-120720-025957},
  urldate = {2023-08-14}
}

@article{avila-etal-2011,
  title = {The {{Onset}} of {{Turbulence}} in {{Pipe Flow}}},
  author = {Avila, K. and Moxey, D. and {de Lozar}, A. and Avila, M. and Barkley, D. and Hof, B.},
  year = 2011,
  month = jul,
  journal = {Science},
  volume = {333},
  number = {6039},
  pages = {192--196},
  doi = {10.1126/science.1203223},
  urldate = {2024-10-24}
}

@article{avila-hof-2013,
  title = {Nature of Laminar-Turbulence Intermittency in Shear Flows},
  author = {Avila, M. and Hof, B.},
  year = 2013,
  month = jun,
  journal = {Phys. Rev. E},
  volume = {87},
  number = {6},
  pages = {063012},
  doi = {10.1103/PhysRevE.87.063012},
  urldate = {2024-10-09}
}

@article{avila-willis-hof-2010,
  title = {On the Transient Nature of Localized Pipe Flow Turbulence},
  author = {Avila, M. and Willis, A.P. and Hof, B.},
  year = 2010,
  month = mar,
  journal = {J. Fluid Mech.},
  volume = {646},
  pages = {127--136},
  issn = {1469-7645, 0022-1120},
  doi = {10.1017/S0022112009993296},
  urldate = {2024-10-30},
  langid = {english}
}

@article{barkley-etal-2015,
  title = {The Rise of Fully Turbulent Flow},
  author = {Barkley, D. and Song, B. and Mukund, V. and Lemoult, G. and Avila, M. and Hof, B.},
  year = 2015,
  month = oct,
  journal = {Nature},
  volume = {526},
  number = {7574},
  pages = {550--553},
  publisher = {Nature Publishing Group},
  issn = {1476-4687},
  doi = {10.1038/nature15701},
  urldate = {2026-06-02},
  copyright = {2015 Springer Nature Limited},
  langid = {english}
}

@article{berizzi-etal-2025a,
  title = {Aerodynamic Performance of a Transonic Airfoil with Spanwise Forcing},
  author = {Berizzi, N. and Gatti, D. and Soldati, G. and Pirozzoli, S. and Quadrio, M.},
  year = 2025,
  month = may,
  journal = {J. Fluid Mech.},
  volume = {1010},
  pages = {A18},
  issn = {0022-1120, 1469-7645},
  doi = {10.1017/jfm.2025.332},
  urldate = {2025-05-09},
  langid = {english}
}

@article{chate-manneville-1988,
  title = {Spatio-Temporal Intermittency in Coupled Map Lattices},
  author = {Chat{\'e}, H. and Manneville, P.},
  year = 1988,
  month = dec,
  journal = {Physica D},
  volume = {32},
  number = {3},
  pages = {409--422},
  issn = {0167-2789},
  doi = {10.1016/0167-2789(88)90065-6},
  urldate = {2025-01-16}
}

@article{chin-etal-2010,
  title = {The Influence of Pipe Length on Turbulence Statistics Computed from Direct Numerical Simulation Data},
  author = {Chin, C. and Ooi, A.S.H. and Marusic, I. and Blackburn, H.M.},
  year = 2010,
  journal = {Phys. Fluids},
  volume = {22},
  number = {115107},
  pages = {1--10}
}

@article{ciliberto-bigazzi-1988,
  title = {Spatiotemporal {{Intermittency}} in {{Rayleigh-B\'enard Convection}}},
  author = {Ciliberto, S. and Bigazzi, P.},
  year = 1988,
  month = jan,
  journal = {Phys. Rev. Lett.},
  volume = {60},
  number = {4},
  pages = {286--289},
  doi = {10.1103/PhysRevLett.60.286},
  urldate = {2025-01-16}
}

@article{coxe-peet-adrian-2022,
  title = {On {{Stokes}}' Second Problem Solutions in Cylindrical and {{Cartesian}} Domains},
  author = {Coxe, D. and Peet, Y. and Adrian, R.},
  year = 2022,
  month = oct,
  journal = {Phys. Fluids},
  volume = {34},
  number = {10},
  pages = {103615},
  issn = {1070-6631},
  doi = {10.1063/5.0118838},
  urldate = {2023-01-25}
}

@article{delozar-hof-2009,
  title = {An {{Experimental Study}} of the {{Decay}} of {{Turbulent Puffs}} in {{Pipe Flow}}},
  author = {{de Lozar}, A. and Hof, B.},
  year = 2009,
  journal = {Phil. Trans. R. Soc. A},
  volume = {367},
  number = {1888},
  eprint = {40485460},
  eprinttype = {jstor},
  pages = {589--599},
  issn = {1364-503X},
  urldate = {2024-10-31}
}

@article{faisst-eckhardt-2004,
  title = {Sensitive Dependence on Initial Conditions in Transition to Turbulence in Pipe Flow},
  author = {Faisst, H. and Eckhardt, B.},
  year = 2004,
  journal = {J. Fluid Mech.},
  volume = {504},
  pages = {343--352}
}

@inproceedings{gallorini-etal-2024,
  title = {The State of Turbulence in a Pipe Flow with Drag Reduction},
  booktitle = {European {{Drag Reduction}} and {{Flow Control Meeting}}},
  author = {Gallorini, E. and Massaro, D. and Schlatter, P. and Quadrio, M.},
  year = 2024,
  address = {Torino (Italy)}
}

@article{gallorini-quadrio-2024,
  title = {Spatial Discretization Effects in Spanwise Forcing for Turbulent Drag Reduction},
  author = {Gallorini, E. and Quadrio, M.},
  year = 2024,
  month = mar,
  journal = {J. Fluid Mech.},
  volume = {982},
  pages = {A11},
  issn = {0022-1120, 1469-7645},
  doi = {10.1017/jfm.2024.107},
  urldate = {2024-03-03},
  langid = {english}
}

@article{gallorini-quadrio-gatti-2022,
  title = {Coherent Near-Wall Structures and Drag Reduction by Spanwise Forcing},
  author = {Gallorini, E. and Quadrio, M. and Gatti, D.},
  year = 2022,
  month = nov,
  journal = {Phys. Rev. Fluids},
  volume = {7},
  number = {11},
  pages = {114602},
  doi = {10.1103/PhysRevFluids.7.114602},
  urldate = {2022-11-22}
}

@inproceedings{gattere-etal-2024,
  title = {On the {{Optimal Period}} of {{Spanwise Forcing}} for {{Turbulent Drag Reduction}}},
  booktitle = {European {{Drag Reduction}} and {{Flow Control Meeting}} ({{EDRFCM}})},
  author = {Gattere, F. and Chiarini, A. and Castelletti, M. and Quadrio, M.},
  year = 2024,
  address = {Torino (Italy)}
}

@article{gatti-etal-2025,
  title = {Turbulent Skin-Friction Drag Reduction via Spanwise Forcing at High {{Reynolds}} Number},
  author = {Gatti, D. and Quadrio, M. and Chiarini, A. and Gattere, F. and Pirozzoli, S.},
  year = 2025,
  month = aug,
  journal = {J. Fluid Mech.},
  volume = {1016},
  pages = {A58, pp.1--23},
  issn = {0022-1120, 1469-7645},
  doi = {10.1017/jfm.2025.10412},
  urldate = {2025-09-02},
  langid = {english}
}

@article{gatti-quadrio-2016,
  title = {Reynolds-Number Dependence of Turbulent Skin-Friction Drag Reduction Induced by Spanwise Forcing},
  author = {Gatti, D. and Quadrio, M.},
  year = 2016,
  journal = {J. Fluid Mech.},
  volume = {802},
  pages = {553--58}
}

@article{hasegawa-quadrio-frohnapfel-2014,
  title = {Numerical Simulation of Turbulent Duct Flows at Constant Power Input},
  author = {Hasegawa, Y. and Quadrio, M. and Frohnapfel, B.},
  year = 2014,
  journal = {J. Fluid Mech.},
  volume = {750},
  pages = {191--209}
}

@article{hof-etal-2005,
  title = {Turbulence {{Regeneration}} in {{Pipe Flow}} at {{Moderate Reynolds Numbers}}},
  author = {Hof, B. and {van Doorne}, C. W. H. and Westerweel, J. and Nieuwstadt, F. T. M.},
  year = 2005,
  month = nov,
  journal = {Phys. Rev. Lett.},
  volume = {95},
  number = {21},
  pages = {214502},
  doi = {10.1103/PhysRevLett.95.214502},
  urldate = {2024-10-30}
}

@article{hof-etal-2006,
  title = {Finite Lifetime of Turbulence in Shear Flows},
  author = {Hof, B. and Westerveel, J. and Schneider, T. and Eckhardt, B.},
  year = 2006,
  journal = {Nature}
}

@article{hof-etal-2008,
  title = {Repeller or {{Attractor}}? {{Selecting}} the {{Dynamical Model}} for the {{Onset}} of {{Turbulence}} in {{Pipe Flow}}},
  shorttitle = {Repeller or {{Attractor}}?},
  author = {Hof, B. and {de Lozar}, A. and Kuik, D. J. and Westerweel, J.},
  year = 2008,
  month = nov,
  journal = {Phys. Rev. Lett.},
  volume = {101},
  number = {21},
  pages = {214501},
  doi = {10.1103/PhysRevLett.101.214501},
  urldate = {2024-10-31}
}

@article{hof-etal-2010,
  title = {Eliminating {{Turbulence}} in {{Spatially Intermittent Flows}}},
  author = {Hof, B. and {de Lozar}, A. and Avila, M. and Tu, X. and Schneider, T. M.},
  year = 2010,
  month = mar,
  journal = {Science},
  volume = {327},
  number = {5972},
  pages = {1491--1494},
  doi = {10.1126/science.1186091},
  urldate = {2024-10-09}
}

@article{jeong-hussain-1995,
  title = {On the Identification of a Vortex},
  author = {Jeong, J. and Hussain, F.},
  year = 1995,
  journal = {J. Fluid Mech.},
  volume = {285},
  pages = {69--94},
  catalog = {0051}
}

@article{kim-adrian-1999,
  title = {Very Large-Scale Motion in the Outer Layer},
  author = {Kim, K. C. and Adrian, R. J.},
  year = 1999,
  month = feb,
  journal = {Phys. Fluids},
  volume = {11},
  number = {2},
  pages = {417--422},
  issn = {1070-6631, 1089-7666},
  doi = {10.1063/1.869889},
  urldate = {2021-12-21},
  langid = {english}
}

@article{lewis-bellan-1990,
  title = {Physical Constraints on the Coefficients of {{Fourier}} Expansion in Cylindrical Coordinates},
  author = {Lewis, H.R. and Bellan, P.M.},
  year = 1990,
  journal = {J. Math. Phys.},
  volume = {31},
  number = {11},
  pages = {2592--2596}
}

@phdthesis{lindgren-1957,
  title = {The Transition Process and Other Phenomena in Viscous Flow},
  author = {Lindgren, R.},
  year = 1957,
  school = {KTH Royal Institute of Technology}
}

@article{liu-etal-2022,
  title = {Turbulence Suppression by Streamwise-Varying Wall Rotation in Pipe Flow},
  author = {Liu, X. and Zhu, H. and Bao, Y. and Zhou, D. and Han, Z.},
  year = 2022,
  journal = {J. Fluid Mech.},
  volume = {951},
  pages = {A35},
  urldate = {2021-01-07}
}

@misc{luchini-2020,
  title = {{{CPL}}},
  author = {Luchini, P.},
  year = 2020,
  howpublished = {Available at https://CPLcode.net}
}

@misc{luchini-2021,
  title = {Introducing {{CPL}}},
  author = {Luchini, P.},
  year = 2021,
  month = nov,
  number = {2012.12143v2},
  eprint = {2012.12143v2},
  publisher = {arXiv},
  urldate = {2022-10-05},
  archiveprefix = {arXiv}
}

@article{luchini-quadrio-2006,
  title = {A Low-Cost Parallel Implementation of Direct Numerical Simulation of Wall Turbulence},
  author = {Luchini, P. and Quadrio, M.},
  year = 2006,
  journal = {J. Comp. Phys.},
  volume = {211},
  number = {2},
  pages = {551--571}
}

@article{massaro-etal-2023,
  title = {Linear Stability of {{Poiseuille}} Flow over a Steady Spanwise {{Stokes}} Layer},
  author = {Massaro, D. and Martinelli, F. and Schmid, P.J. and Quadrio, M.},
  year = 2023,
  month = oct,
  journal = {Phys. Rev. Fluids},
  volume = {8},
  number = {10},
  pages = {103902},
  doi = {10.1103/PhysRevFluids.8.103902},
  urldate = {2024-11-17}
}

@article{meseguer-trefethen-2003,
  title = {Linearized Pipe Flow to {{Reynolds}} Number 10\textasciicircum 7},
  author = {Meseguer, {\'A}. and Trefethen, L.N.},
  year = 2003,
  month = mar,
  journal = {J. Comp. Phys.},
  volume = {186},
  number = {1},
  pages = {178--197},
  issn = {00219991},
  doi = {10.1016/S0021-9991(03)00029-9},
  urldate = {2024-11-16},
  copyright = {https://www.elsevier.com/tdm/userlicense/1.0/},
  langid = {english}
}

@article{moxey-barkley-2010,
  title = {Distinct Large-Scale Turbulent-Laminar States in Transitional Pipe Flow},
  author = {Moxey, D. and Barkley, D.},
  year = 2010,
  month = may,
  journal = {PNAS},
  volume = {107},
  number = {18},
  pages = {8091--8096},
  doi = {10.1073/pnas.0909560107},
  urldate = {2024-06-23}
}

@article{mukund-hof-2018,
  title = {The Critical Point of the Transition to Turbulence in Pipe Flow},
  author = {Mukund, V. and Hof, B.},
  year = 2018,
  month = mar,
  journal = {J. Fluid Mech.},
  volume = {839},
  pages = {76--94},
  issn = {0022-1120, 1469-7645},
  doi = {10.1017/jfm.2017.923},
  urldate = {2023-06-16},
  langid = {english}
}

@article{peixinho-mullin-2006,
  title = {Decay of {{Turbulence}} in {{Pipe Flow}}},
  author = {Peixinho, J. and Mullin, T.},
  year = 2006,
  month = mar,
  journal = {Phys. Rev. Lett.},
  volume = {96},
  number = {9},
  pages = {094501},
  doi = {10.1103/PhysRevLett.96.094501},
  urldate = {2024-10-30}
}

@article{quadrio-frohnapfel-hasegawa-2016,
  title = {Does the Choice of the Forcing Term Affect Flow Statistics in {{DNS}} of Turbulent Channel Flow?},
  author = {Quadrio, M. and Frohnapfel, B. and Hasegawa, Y.},
  year = 2016,
  journal = {Eur. J. Mech. B / Fluids},
  volume = {55},
  pages = {286--293}
}

@article{quadrio-ricco-2011,
  title = {The Laminar Generalized {{Stokes}} Layer and Turbulent Drag Reduction},
  author = {Quadrio, M. and Ricco, P.},
  year = 2011,
  journal = {J. Fluid Mech.},
  volume = {667},
  pages = {135--157},
  catalog = {00429},
  myfile = {file:///home/mq/Library/quadrio-ricco-2011.pdf}
}

@article{quadrio-ricco-viotti-2009,
  title = {Streamwise-Traveling Waves of Spanwise Wall Velocity for Turbulent Drag Reduction},
  author = {Quadrio, M. and Ricco, P. and Viotti, C.},
  year = 2009,
  journal = {J. Fluid Mech.},
  volume = {627},
  pages = {161--178},
  doi = {10.1017/S0022112009006077}
}

@article{quadrio-sibilla-2000,
  title = {Numerical Simulation of Turbulent Flow in a Pipe Oscillating around Its Axis},
  author = {Quadrio, M. and Sibilla, S.},
  year = 2000,
  journal = {J. Fluid Mech.},
  volume = {424},
  pages = {217--241}
}

@article{rai-moin-1991,
  title = {Direct Simulations of Turbulent Flow Using Finite-Difference Schemes},
  author = {Rai, M.M. and Moin, P.},
  year = 1991,
  journal = {J. Comp. Phys.},
  volume = {96},
  pages = {15}
}

@article{reynolds-1883,
  title = {An Experimental Investigation on the Circumstances Which Determine Whether the Motion of Water Shall Be Direct or Sinuous, and the Law of Resistance in Parallel Channels},
  author = {Reynolds, O.},
  year = 1883,
  journal = {Proc. R. Soc. Lond. A},
  volume = {35},
  pages = {84}
}

@article{ricco-etal-2012,
  title = {Changes in Turbulent Dissipation in a Channel Flow with Oscillating Walls},
  author = {Ricco, P. and Ottonelli, C. and Hasegawa, Y. and Quadrio, M.},
  year = 2012,
  journal = {J. Fluid Mech.},
  volume = {700},
  pages = {77--104}
}

@article{ricco-quadrio-2008,
  title = {Wall-Oscillation Conditions for Drag Reduction in Turbulent Channel Flow},
  author = {Ricco, P. and Quadrio, M.},
  year = 2008,
  journal = {Int. J. Heat Fluid Flow},
  volume = {29},
  pages = {601--612}
}

@article{ricco-skote-leschziner-2021,
  title = {A Review of Turbulent Skin-Friction Drag Reduction by near-Wall Transverse Forcing},
  author = {Ricco, P. and Skote, M. and Leschziner, M. A.},
  year = 2021,
  journal = {Prog. Aero. Sci.},
  volume = {123},
  pages = {100713},
  issn = {0376-0421},
  urldate = {2021-06-06},
  langid = {english}
}

@article{ruby-foysi-2025,
  title = {Effects by Oscillation Control and Compressibility in Supersonic Turbulent Channel Flow},
  author = {Ruby, M. and Foysi, H.},
  year = 2025,
  month = aug,
  journal = {Int. J. Heat Fluid Flow},
  volume = {114},
  pages = {109820},
  issn = {0142-727X},
  doi = {10.1016/j.ijheatfluidflow.2025.109820},
  urldate = {2025-04-23}
}

@article{russo-luchini-2017,
  title = {A Fast Algorithm for the Estimation of Statistical Error in {{DNS}} (or Experimental) Time Averages},
  author = {Russo, S. and Luchini, P.},
  year = 2017,
  journal = {J. Comput. Phys.},
  volume = {347},
  pages = {328--340},
  issn = {0021-9991},
  urldate = {2021-10-31},
  langid = {english}
}

@book{schmid-henningson-2001,
  title = {Stability and {{Transition}} in {{Shear Flows}}},
  author = {Schmid, P.J. and Henningson, D.S.},
  year = 2001,
  publisher = {Springer}
}

@article{takeishi-etal-2015,
  title = {Localized Turbulence Structures in Transitional Rectangular-Duct Flow},
  author = {Takeishi, K. and Kawahara, G. and Wakabayashi, H. and Uhlmann, M. and Pinelli, A.},
  year = 2015,
  month = nov,
  journal = {J. Fluid Mech.},
  volume = {782},
  pages = {368--379},
  issn = {0022-1120, 1469-7645},
  doi = {10.1017/jfm.2015.546},
  urldate = {2025-07-09},
  langid = {english}
}

@article{tuckerman-chantry-barkley-2020,
  title = {Patterns in {{Wall-Bounded Shear Flows}}},
  author = {Tuckerman, L. and Chantry, M. and Barkley, D.},
  year = 2020,
  month = jan,
  journal = {Annual Review of Fluid Mechanics},
  volume = {52},
  number = {Volume 52, 2020},
  pages = {343--367},
  publisher = {Annual Reviews},
  issn = {0066-4189, 1545-4479},
  doi = {10.1146/annurev-fluid-010719-060221},
  urldate = {2026-06-02},
  langid = {english}
}

@article{vandoorne-westerweel-2008,
  title = {The Flow Structure of a Puff},
  author = {{van Doorne}, C.W.H. and Westerweel, J.},
  year = 2008,
  month = nov,
  journal = {Phil. Trans. R. Soc. A},
  volume = {367},
  number = {1888},
  pages = {489--507},
  doi = {10.1098/rsta.2008.0227},
  urldate = {2024-11-16}
}

@article{willis-kerswell-2007,
  title = {Critical Behavior in the Relaminarization of Localized Turbulence in Pipe Flow},
  author = {Willis, A.P. and Kerswell, R.R.},
  year = 2007,
  month = jan,
  journal = {Phys. Rev. Lett.},
  volume = {98},
  number = {1},
  pages = {014501},
  doi = {10.1103/PhysRevLett.98.014501},
  urldate = {2024-10-30}
}

@article{wu-moin-2008,
  title = {A Direct Numerical Simulation Study of the Mean Velocity Characteristics in Turbulent Pipe Flow},
  author = {Wu, X. and Moin, P.},
  year = 2008,
  journal = {J. Fluid Mech.},
  volume = {608},
  pages = {81--112}
}

@article{wygnanski-champagne-1973,
  title = {On Transition in a Pipe. {{Part}} 1. {{The}} Origin of Puffs and Slugs and the Flow in a Turbulent Slug},
  author = {Wygnanski, I. J. and Champagne, F. H.},
  year = 1973,
  month = jun,
  journal = {J. Fluid Mech.},
  volume = {59},
  number = {2},
  pages = {281--335},
  issn = {1469-7645, 0022-1120},
  doi = {10.1017/S0022112073001576},
  urldate = {2024-10-30},
  langid = {english}
}

@article{wygnanski-sokolov-friedman-1975,
  title = {On Transition in a Pipe. {{Part}} 2. {{The}} Equilibrium Puff},
  author = {Wygnanski, I. and Sokolov, M. and Friedman, D.},
  year = 1975,
  month = may,
  journal = {J. Fluid Mech.},
  volume = {69},
  number = {2},
  pages = {283--304},
  issn = {1469-7645, 0022-1120},
  doi = {10.1017/S0022112075001449},
  urldate = {2024-10-30},
  langid = {english}
}

@article{xiao-etal-2024,
  title = {Direct numerical simulation of drag reduction in rotating pipe flow up to ${R}e_\tau \approx 3000$},
  author = {Xiao, M. and Ceci, A. and Orlandi, P. and Pirozzoli, S.},
  year = 2024,
  month = oct,
  journal = {J. Fluid Mech.},
  volume = {996},
  pages = {A24},
  issn = {0022-1120, 1469-7645},
  doi = {10.1017/jfm.2024.811},
  urldate = {2026-01-21},
  langid = {english}
}

\end{document}